\documentclass[journal,10pt]{IEEEtran}

\makeatletter
\def\ps@headings{%
\def\@oddhead{\mbox{}\scriptsize\rightmark \hfil \thepage}%
\def\@evenhead{\scriptsize\thepage \hfil \leftmark\mbox{}}%
\def\@oddfoot{}%
\def\@evenfoot{}}
\makeatother
\ifCLASSINFOpdf
\else
\fi

\newcounter{problem}
\newcounter{save@equation}
\newcounter{save@problem}

\makeatletter

\usepackage{epsfig}
\usepackage{array}
\usepackage[short]{optidef}
\usepackage{amsthm}
\usepackage{mathrsfs}
\usepackage{flexisym}
\usepackage{algorithmicx}
\usepackage{algorithm}
\usepackage{siunitx}
\usepackage{comment}
\usepackage{soul}
\usepackage{algpseudocode}
\usepackage{cite}
\usepackage{url}
\usepackage[utf8]{inputenc}
\usepackage[T1]{fontenc}
\usepackage{amsmath}
\usepackage{amsfonts}
\usepackage{amssymb}
\usepackage[version=4]{mhchem}
\usepackage{stmaryrd}
\usepackage{graphicx}
\usepackage[export]{adjustbox}
\usepackage{breqn}
\usepackage{lipsum}
\usepackage{mathtools}
\usepackage{caption}
\usepackage{float}
\usepackage{tikz}
\usepackage{subcaption}
\usepackage{soul}
\usepackage{blindtext, graphicx}
\usepackage{cuted}
\usepackage[hidelinks]{hyperref}
\usepackage{color}
\usepackage{multicol}

\newcolumntype{L}[1]{>{\raggedright\let\newline\\\arraybackslash\hspace{0pt}}m{#1}}
\newcolumntype{C}[1]{>{\centering\let\newline\\\arraybackslash\hspace{0pt}}m{#1}}
\newcolumntype{R}[1]{>{\raggedleft\let\newline\\\arraybackslash\hspace{0pt}}m{#1}}

\newtheoremstyle{case}{}{}{}{}{}{:}{ }{}
\newcommand{\bc}{\begin{center}}
\newcommand{\ec}{\end{center}}
\newcommand{\be}{\begin{equation}}
\newcommand{\ee}{\end{equation}}

\newcommand{\bnu}{\begin{enumerate}}
\newcommand{\enu}{\end{enumerate}}

\begin{document}

\title{Efficient VoIP Communications through LLM-based Real-Time Speech Reconstruction and Call Prioritization for Emergency Services}
%\title{Enhancing Emergency Service Efficiency through LLM-based Real-Time Speech Reconstruction and Call Prioritization in Bandwidth Constrained VOIP Networks}
\author{Danush Venkateshperumal, Rahman Abdul Rafi, Shakil Ahmed,~\IEEEmembership{Member,~IEEE}, and Ashfaq Khokhar,~\IEEEmembership{Fellow,~IEEE}

Corresponding Author: Shakil Ahmed (email: shakil@iastate.edu)

\vspace*{-0.5 cm}
\thanks{Corresponding Author: Shakil Ahmed (email: shakil@iastate.edu)

D. Venkateshperumal, R. A. Rafi, S. Ahmed, and A. Khokhar are with the Department of Electrical and Computer Engineering, Iowa State University, Ames, Iowa, USA (email: \{da2016, rah7, shakil, ashfaq\}@iastate.edu). }
}

% Make the title area
\maketitle

\begin{abstract}
Emergency communications systems often face significant disruptions between victims and dispatchers due to packet loss, bandwidth constraints, poor signal quality, delay, and jitter in VoIP systems, leading to degraded quality of service in real-time and eventually making it challenging for dispatchers to interpret the urgency and details of a situation accurately. Additionally, victims in distress may struggle to convey critical information due to speech disorders, panic, and background noise, further exacerbating these challenges. This results in incomplete or ambiguous communication, hindering dispatchers' ability to assess situations accurately and, when combined with staffing shortages in many emergency response centers, causing delays in coordination and the timely delivery of emergency assistance. This paper proposes solving the abovementioned issues by leveraging Large Language Models (LLMs) to improve emergency response efficiency. This is done by reconstructing incomplete speech, filling contextual gaps to provide meaningful outputs, and prioritizing the emergency calls accordingly through a multi-step process involving real-time data processing and advanced severity, rule-based and keyword classification techniques.. The system integrates real-time audio transcription and uses a Retrieval-Augmented Generation (RAG) model to generate relevant responses. This output is further used to classify the severity of the emergency to ensure that critical cases are prioritized using\texttt{ Twilio }and\texttt{ AssmbleAI }Application Programming Interfaces (APIs). Evaluation of the proposed model demonstrates high Conceptual Precision (100\%) and favorable Bilingual Evaluation Understudy (BLEU) and Recall-Oriented Understudy for Gisting Evaluation (ROUGE) scores across diverse emergency scenarios, indicating robust alignment with real-world needs\footnote{The code of paper is available at \cite{Venkateshperumal2024RAG}}. These results highlight the system’s potential to improve emergency communication efficiency, optimize response workflows, and prioritize critical cases effectively.
\end{abstract}

\section{Introduction} 
\IEEEPARstart{M}{ore} than 240  million emergency calls are placed in the U.S. annually, and over half of the emergency centers within the U.S. face a staffing crisis, particularly during peak times within major cities \cite{nena_stats}. The problem becomes severe in poor networking conditions, where connectivity issues compound the risk of miscommunication \cite{vriezekolk2016assessing}. 
%Staffing shortages significantly impact the efficiency of handling emergency calls, particularly during crises, leading to prolonged wait times and inefficient resource allocation, especially in areas with limited network connectivity. 
\textcolor{black}{Such conditions degrade audio quality, resulting in fragmented and unintelligible communications that hinder dispatchers' ability to assess and respond to emergencies accurately.} Consequently, connectivity and staffing issues can lead to significant delays in emergency response, resource deployment, and the ability to address each call with the same level of urgency, potentially causing life-threatening situations \cite{911Survey, IAEDPressRelease, neusteter2019911}.
\textcolor{black}{Studies show the critical impact of response times on emergency mortality rates. For instance, each minute of delay in response time increases the mortality rate by 1\% in critical trauma cases \cite{alshammari2024impact}. Additionally, delays of five minutes or more in notifying an ambulance can significantly increase fatality rates, up to 30\% \cite{brodsky1990emergency}. These statistics underscore the importance of efficient and timely communication in emergency services.}

Emergency communication systems rely heavily on voice communication via Internet protocols (VoIP). When victims dial the emergency number, dispatchers at a local emergency center manually assess the emergency details to respond. 
%However, in some regions, the systems automatically display the caller's phone number and address, improving response times regardless of whether the caller's voice is fragmented or incomprehensible. 
Nonetheless, these wireless calls may route to incorrect dispatch centers, leading to the caller's location or number being lost, which can result in dangerous delays, especially if communication fails or the call drops \cite{wells2020emergency}.
VoIP 911 calls may not connect to the 911 call center serving the caller's current location. In some cases, they may be routed to the administrative line of a 911 call center, which may not be staffed after hours or handled by trained 911 operators.
Federal Communications Commission (FCC) \cite{fcc_voip_911} warns that even when VoIP 911 calls connect correctly, they may fail to automatically transmit the caller's phone number and location information, making it difficult for dispatchers to locate the victim.
Moreover, VoIP users are often required to manually provide their location information to their VoIP service providers and update it whenever they change locations to ensure proper functionality of the 911 service.
VoIP services may become inoperable during power outages or internet failures, including when the network is congested, or bandwidth is constrained, further exacerbating delays in emergency response.

Packet loss occurs when data packets fail to reach their destination, significantly impacting VoIP call quality. While small amounts of packet loss might not noticeably affect communication, higher rates can severely degrade audio quality. Maintaining packet loss below 1\% is crucial to achieve voice quality comparable to traditional  Public Switched Telephone Network (PSTN) services \cite{james2004implementing}. Loss rates between 5\% and 10\% can significantly affect call quality \cite{mansfield2009computer}. Moreover, burst packet losses where several consecutive packets are lost, can lead to noticeable gaps in audio, exacerbating the the degradation of call quality \cite{sun2006voice,boutremans2002impact}.
Bandwidth limitations, especially during periods of high network usage, can restrict the data capacity available for real-time audio transmission, causing delays or dropped audio packets. These constraints directly impact voice quality and can result in degraded call clarity due to reduced packet flow \cite{manousos2005voice, james2004implementing}.
Unreliable or weak network connections contribute to degraded audio quality, leading to interruptions and call dropouts \cite{boutremans2002impact}.
Environmental noise, such as background sounds or callers speaking softly to avoid detection, poses challenges to communication clarity. Background noise significantly affects voice quality, especially when combined with packet losses \cite{manousos2005voice}. Additionally, using silence suppression in noisy conditions can cause ``noise-pumping'' effects, further degrading voice quality \cite{james2004implementing}.
Delay refers to the time it takes for packets to reach their destination, while jitter is the variation in packet arrival times. Without effective buffering, these issues can lead to choppy or out-of-order audio, severely disrupting the flow of conversation \cite{sun2006voice}. Network failures can increase delay and jitter, negatively impacting call quality \cite{boutremans2002impact}.

\textcolor{black}{When responding to multiple emergency calls simultaneously, dispatchers often struggle to prioritize one case over another with appropriate urgency. This challenge is exacerbated by bandwidth constraints, VoIP limitations, and random and bursty packet loss, which degrade audio quality and make it difficult to assess each situation's severity accurately. Additionally, poor signal quality and lack of communications interoperability, particularly involving critical communications equipment, can lead to severe miscommunications or delays. Dispatchers frequently transfer calls between various departments multiple times to ensure calls reach the appropriate emergency responders. This repeated transferring further increases the risk of miscommunications and delays, potentially causing life-threatening consequences \cite{burroughs2017three}.}
The mortality rates in medical emergencies rise sharply when response times exceed five minutes \cite{blackwell2002response}. Hence, it is essential to prevent scenarios involving degraded communication quality and ensure that cases are quickly and accurately classified based on urgency, all while providing clear context without causing any additional delay.

\textcolor{black}{Many local call centers operate independently, affecting not only call centers but also emergency medical services (EMS), particularly in ambulance allocation to emergency departments (EDs) \cite{acuna2020ambulance}.} This makes it hard to aggregate data and coordinate responses to the region's victims \cite{neusteter2019911}. There are also significant challenges in ensuring that emergency communication systems are staffed with qualified personnel who receive proper training to handle high-stress situations effectively \cite{eslami2024covid}. 
\textcolor{black}{The lack of coordination in services like EMS frequently results in significant delays. Inefficient data handling contributes to ED overcrowding}, increasing mortality rates, decreased quality of care, longer waiting times, and ambulance offload delays \cite{acuna2020ambulance}. 
These limitations require a need to develop a system to help coordinate communication and resource prioritization better to improve emergency response efficiency. Furthermore, by dynamically prioritizing cases based on their criticality, life-threatening emergencies receive immediate attention while less urgent situations are appropriately scheduled.

The paper proposes a system that leverages a LLM to address communication challenges, allowing dispatchers to focus on immediate responses.
LLMs present a transformative solution to the persistent communication challenges in emergency response systems. By leveraging their capability to interpret and generate language accurately, LLMs can help bridge communication gaps, especially under poor networking conditions or when emergency calls are fragmented or unclear \textcolor{black}{due to packet loss. The model} can process voice-to-text transcriptions, predict missing details, and provide real-time contextual assistance, allowing dispatchers to focus on immediate priorities without re-querying callers repeatedly. LLMs excel in dynamically prioritizing critical information and guiding dispatchers through decision-making, making them invaluable in ensuring that life-threatening situations receive immediate attention. By streamlining call handling across decentralized centers and supporting dispatcher accuracy, LLMs provide an efficient, scalable approach to optimizing emergency response workflows, enhancing response times, and mitigating delays in critical scenarios.

Our proposed LLM-based system is designed to integrate with existing emergency communication infrastructure, automatically reconstructing fragmented or unclear calls to ensure dispatchers receive coherent and complete information promptly. Additionally, the system incorporates a separate severity classification module to assess the reconstructed information and prioritize cases based on their criticality. This approach aims to improve the accuracy of information received by dispatchers to ensure that life-threatening emergencies are addressed immediately while less-urgent emergencies may be addressed accordingly. By streamlining communication and prioritization processes, the proposed system has the potential to enhance the overall effectiveness and responsiveness of emergency services.
Our proposed model surpasses traditional, manual methods for handling emergency calls by reducing delays, errors, and inconsistencies in evaluating the urgency of each situation. This advantage is particularly evident during high-stress incidents, where callers may struggle to convey their needs clearly, and the full scope of the emergency may not be immediately obvious. Our model addresses these challenges by introducing an LLM to:

\textit{Increase communication clarity:} 
\textcolor{black}{In case of unclear or fragmented speech caused due to various human, device, and/or network factors including panic, speech disorders, broken signals, packet loss (random and bursty), and bandwidth limitations, our LLM model can reconstruct the audio so that the broken speech would be understandable. This is achieved by breaking down each word within the input audio transcription through tokenization, allowing the model to analyze and process individual components effectively. Utilizing context comprehension, the LLM infers missing or unclear parts of the conversation, ensuring that the reconstructed message is both coherent and actionable for dispatchers. This way, we can enhance the accuracy of emergency assessments, facilitating more timely and effective responses.}

\textit{Efficient Prioritization of Emergencies:}
\textcolor{black}{During peak hours, managing multiple emergency calls presents significant challenges, especially in prioritizing emergencies based on their severity. Improper prioritization can be life-threatening for victims in high-risk situations if they do not receive prompt assistance over lower-risk cases. Our system addresses this issue by prioritizing emergencies according to the caller's limited context, which is achieved by implementing a rule-based severity classification mechanism integrated with LLM and a Retrieval-Augmented Generation (RAG) model.}

\textit{Addressing Language Barriers and Accessibility:}
\textcolor{black}{Language barriers significantly impact emergency communications, often leading to delays, misunderstandings, and increased stress for both victims and dispatchers \cite{meischke2010emergency}. Callers with Limited English Proficiency (LEP) frequently struggle to convey critical information, causing considerable delays in emergency response efforts. Although over-the-phone interpreter services are available to facilitate communication between dispatchers and callers, these services may not be suitable for high-stress situations \cite{carroll2013serving}. Our system addresses this challenge by leveraging advanced LLM and natural language processing (NLP) techniques to assist dispatchers in understanding callers, even when language barriers exist. %The system enhances communication clarity, reduces response times, and improves overall emergency response effectiveness by reconstructing fragmented speech, performing intent recognition, and providing contextual interpretations.
}

\textit{Reducing Dependence on Manual Processes:}
Without adequate guidelines, reliance on manual processes within EMS increases the risk of human errors \cite{poranen2024human}. These manual processes may affect situational awareness, a crucial component for effective decision-making in \textcolor{black}{Emergency Medical Dispatch (EMD)} \cite{blandford2004situation}.
\textcolor{black}{Our model addresses these challenges by reconstructing fragmented and unclear speech into coherent and understandable outputs using advanced NLP techniques. This allows dispatchers to quickly understand the caller's message without manually interpreting or transcribing broken speech, reducing human error risk. The proposed system enhances situational awareness in EMD, enabling dispatchers to make informed decisions rapidly and efficiently.}

\subsection{Related Works}
Authors in \cite{powers2023using} introduced LLM-based systems to process and categorize disaster communications using social media during Hurricane Harvey. This work illustrated how large pre-trained language models could enhance disaster communications by automatically detecting urgent messages. Their results demonstrated the efficacy of LLMs by identifying life-threatening situations in real-time, particularly during overloaded traditional communications systems. Similar works had been done by \cite{sufi2022automated} and \cite{madichetty2021stacked}. However, those contributions are limited to social media platforms only. 
The authors in \cite{otal2024llm} built LLM platform for public collaboration in \cite{otal2024llm}, where they explored the integration of open-source LLMs, specifically \texttt{Llama2}, into emergency response frameworks. Their research demonstrated the potential of LLMs in assisting emergency dispatchers by analyzing emergency calls in real-time, providing relevant instructions, and guiding the public during large-scale crises through social media-informed emergency response systems. Their study showed that LLMs could bridge linguistic gaps in diverse urban centers and improve situational awareness by processing unstructured data from various sources during critical incidents. However, their study did not address the situation in poor networking conditions. 

In a related study by \cite{chin2021early}, an AI model was developed to assess the emotional state of callers during out-of-hospital cardiac arrest emergency dispatch communications. This model utilized mel-frequency cepstral coefficients to analyze audio recordings, which capture emotional cues from the caller's voice. The AI system demonstrated the potential to pre-screen emotionally stable callers, allowing dispatchers to focus on cases involving emotionally unstable individuals. This approach proved effective in the early recognition of emotional distress, particularly within the first 10 seconds of the call. Similar work by \cite{blushtein2020identifying} used call-tracking technology to reduce the abuse of calls in emergency services, improving efficiency by automatically diverting non-emergency calls without missing real emergencies.

A recent study by \cite{ma2024leveraging} enhanced \textcolor{black}{Speech Emotion Recognition (SER)} using data2vec, GPT-4, and \textcolor{black}{Azure Text-To-Speech (TTS)} to generate high-quality synthetic emotional speech data. Similar works on \textcolor{black}{SER} have been done by \cite{huang2014speech} and \cite{liu2018speech}. Utilizing speech emotion in this context might help classify the situation's urgency.
 Above mentioned works highlight recent efforts to enhance dispatcher-caller interactions and improve system efficiency. However, research addressing the limitations of poor network conditions and limited stuffing in emergency communication remains limited.

\subsection{Our Contribution}
This paper introduces a system that leverages LLM to improve emergency communications clarity and response efficiency in real time by addressing multiple challenges in these critical moments. Our contributions are summarized as follows:

\subsubsection{Handling fragmented or incoherent speech}
The system reconstructs fragmented or incoherent communication, such as speech impacted by panic, illness, or technical challenges like random and bursty packet loss, bandwidth constraints, and poor signal quality, ensuring dispatchers receive coherent and actionable information without delays during emergencies. To achieve this, the system leverages advanced NLP techniques, including tokenization, contextual analysis, and semantic reconstruction, effectively transforming disjointed speech into meaningful and structured data.

\subsubsection{Contextual gap filling}
The system processes incomplete sentences and generates real-time inferences to relay critical and relevant information to dispatchers, enabling quicker and more accurate responses through advanced NLP techniques.

\subsubsection{Dynamic prioritization of emergency data}
The system classifies the acquired data and infers the criticality of the situation by leveraging a combination of severity classification algorithms and machine learning models trained on historical emergency call data. Specifically, the system analyzes linguistic features, such as keywords, tone, and urgency-related phrases, alongside contextual cues from similar past cases. By applying a weighted scoring mechanism, it calculates a severity score that ranks each case based on urgency, even when communication is affected by random packet loss or bandwidth limitations, to ensure that low-priority cases do not delay handling life-threatening emergencies.
By delivering clearer, structured, and complete information to dispatchers, our system aids in the initial stages of emergency response coordination, improving response times and outcomes.

\subsubsection{Implementation of RAG model}
The system implements a RAG model to generate predictive and contextually relevant responses based on real-time transcription data. The RAG model utilizes a custom \textcolor{black}{Facebook AI Similarity Search (FAISS)} index built from a database of emergency call transcripts to retrieve similar past situations. Leveraging these past scenarios, the model enhances the relevance and accuracy of its generated responses.

\subsubsection{API integrations}
We have implemented the\texttt{ Twilio }Media Stream \textcolor{black}{Application Programming Interface (API)} and \texttt{AssemblyAI} API to enable real-time transcription and analysis of emergency calls. The \texttt{ Twilio }API captures audio from calls in real-time, sending it to our WebSocket server, while AssemblyAI’s API processes this audio into text transcripts.  This seamless integration allows our proposed LLM-based system to receive and process transcriptions in real-time, enabling prompt and accurate analysis for emergency response.

\section{Use Cases}
\begin{figure}[h]
\centering
\includegraphics[width=\linewidth]{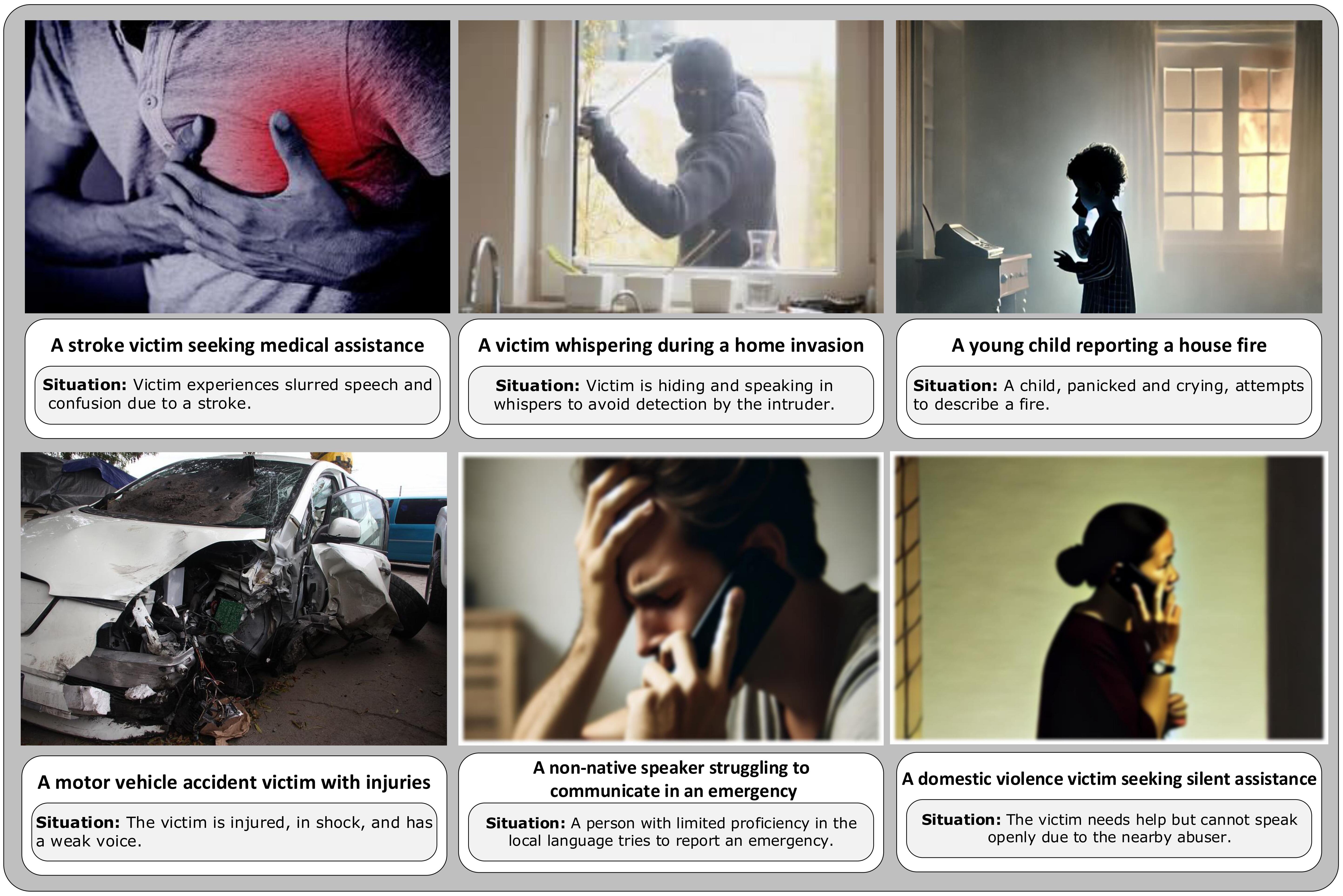}
\caption{Potential Use Cases}
\label{fig:Usecasescenarios}
\end{figure}

In emergencies, effective communications between the victim and the dispatcher is essential. Unclear communications can cause significant delays, potentially slowing the emergency response. Delays, assumptions, and complacency can prove fatal in these situations\cite{gogalniceanu2021capturing}. Unfortunately, many factors may cause such a delay in response, including weak signals, speech problems, vertigo, language barriers, and disturbances of consciousness\cite{handschu2003emergency}. \textcolor{black}{Weak signals, caused by packet loss, bandwidth constraints, and poor signal quality \cite{james2004implementing}, can in turn also lead to delays. }
Regardless of broken speech or muffled or unclear communication, this system ensures that the dispatcher will receive a coherent and actionable message that can be relied upon. This makes the system extremely useful in various cases ranging from natural disasters and criminal incidents to medical cases.
In certain situations, victims may need to whisper to avoid detection. They may also panic while they try to convey their message. This causes a significant shift in their voice, such that it may be fragmented and unperceivable to a regular person. In such cases, the LLM-based system would help understand the fragmented speech and “fill in the gaps” wherever necessary. This ensures that the dispatcher receives the full context of the victim’s speech and can take appropriate action, even in undesirable conditions.
Fig.~\ref{fig:Usecasescenarios} shows various scenarios that may lead victims to call emergency services.

\textit{1. A young child reporting a house fire:}
In a scenario where a child is experiencing a house fire, they would likely be panicked, crying, and facing challenges such as breathlessness from the smoke. When they try to describe a house fire to emergency dispatchers, their speech would be incoherent, and the young child would have a limited vocabulary in general, making it difficult for the child to communicate effectively over the phone. In such a case, this LLM-based model would be helpful to clarify speech affected by panic and tears. The system would interpret the fragmented speech and extract crucial information, such as the fire’s location and severity, based on the limited input from the child.

\textit{2. Emergency situations in remote locations:}
When the victim is situated in a particularly remote area, the communications strength of the network may be unreliable \textcolor{black}{due to factors like limited bandwidth, high latency, packet loss, and poor signal quality.} Here, the system can acquire whatever voice signals are comprehensible and reconstruct an actionable message, regardless of how broken the communications is.

\textit{3. Language barriers:}
A person with limited proficiency in the local language may find it hard to communicate during an emergency. Language barriers with LEP callers negatively impact communications and care outcomes\cite{meischke2010emergency,carroll2013serving}, where the dispatcher may need to ask the victim to repeat their speech multiple times. This would waste essential time. The LLM-based model would understand the vague speech and reconstruct the broken sentences to ensure accurate and efficient communication, minimizing repetition and time wastage.

\textit{4. A Domestic violence victim seeking silent assistance:}
In a situation where a victim needs help but cannot speak openly due to a nearby abuser, they may need to communicate through whispered speech. In such cases, the dispatcher may find it challenging to comprehend the faint speech. Here, we can leverage the system to amplify the soft speech and infer the nature of the speech with minimal verbal cues. This can then be relayed to the dispatcher to take appropriate action.

\textit{5. Medical emergencies:}
Medical emergencies present one of the most significant challenges, where a victim may struggle to communicate with an emergency dispatcher. In these cases, victims may face various symptoms that impair speech, such as panic, pain, confusion, or dizziness. The victim may try to explain their condition, but due to the effects of the medical condition, it may be difficult for them to communicate clearly. The system ensures that crucial information about the victim's condition is conveyed in real-time, which could help improve the situation through basic medical procedures performed by the victim or by guiding medical responders more effectively. 
Considering the case of a stroke victim, their speech is usually heavily impaired. They often suffer from symptoms such as slurred speech, vertigo, and motor deficits\cite{handschu2003emergency}, affecting their ability to communicate their needs effectively. In such cases, time is of the essence. A stroke victim trying to clarify themselves repeatedly wastes essential time, which could lead to permanent disability or even death\cite{lee2021impact}. 

This LLM-based system plays a crucial role by processing the victim’s fragmented speech, inferring the remaining speech based on context, and providing the dispatcher with real-time information about the victim’s condition. For example, a seizure victim attempting to say, “I... shaking... c-can’t... help…” might have their speech inferred by the LLM model as: “I am shaking uncontrollably. I cannot speak clearly. Please help.” This information would allow the dispatcher to facilitate immediate medical assistance for the victim.
By prioritizing critical cases, the system effectively reduces delays in attending to life-threatening emergencies, ultimately improving response times and outcomes.

\section{LLM-based Emergency Services}
The proposed system model addresses the limitations commonly encountered in emergency response scenarios, such as poor audio quality, broken language, and network issues. In this model, a caller (the victim) places a distress call to a dispatcher. Due to potential challenges, such as incomplete language or degraded network conditions, the initial call may need more coherence or clarity. \textcolor{black}{The LLM is utilized within the system} to overcome these limitations. The process begins as follows:

\textit{Call Initiation and Initial Query: }Upon receiving a call, the system initiates a default prompt to the caller, such as, \textit{"What is your emergency?" }This ensures consistency in gathering the caller's response.

\textit{LLM Integration and Real-Time Transcription:} The caller's response, potentially fragmented due to language or network issues, is transcribed in real time and fed into the LLM. The LLM is integrated with a RAG model, allowing it to access and retrieve relevant contextual information to fill in gaps or reconstruct incomplete phrases.
\begin{figure}[htbp]
\centering
\begin{subfigure}{0.5\textwidth}
\centering
\includegraphics[width=\linewidth]{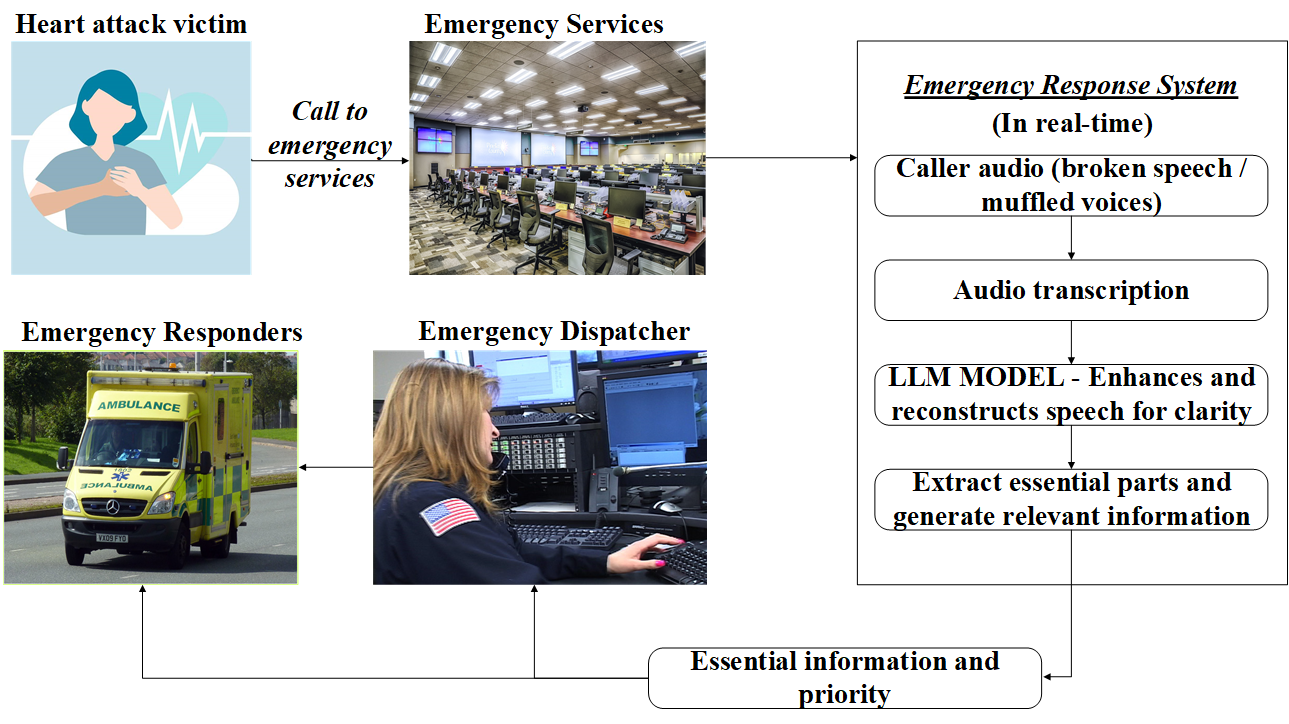}
\caption{LLM-Driven Framework for Emergency Response}
\label{fig:usecase1}
\end{subfigure}
\hfill
\begin{subfigure}{0.5\textwidth}
\centering
\includegraphics[width=\linewidth]{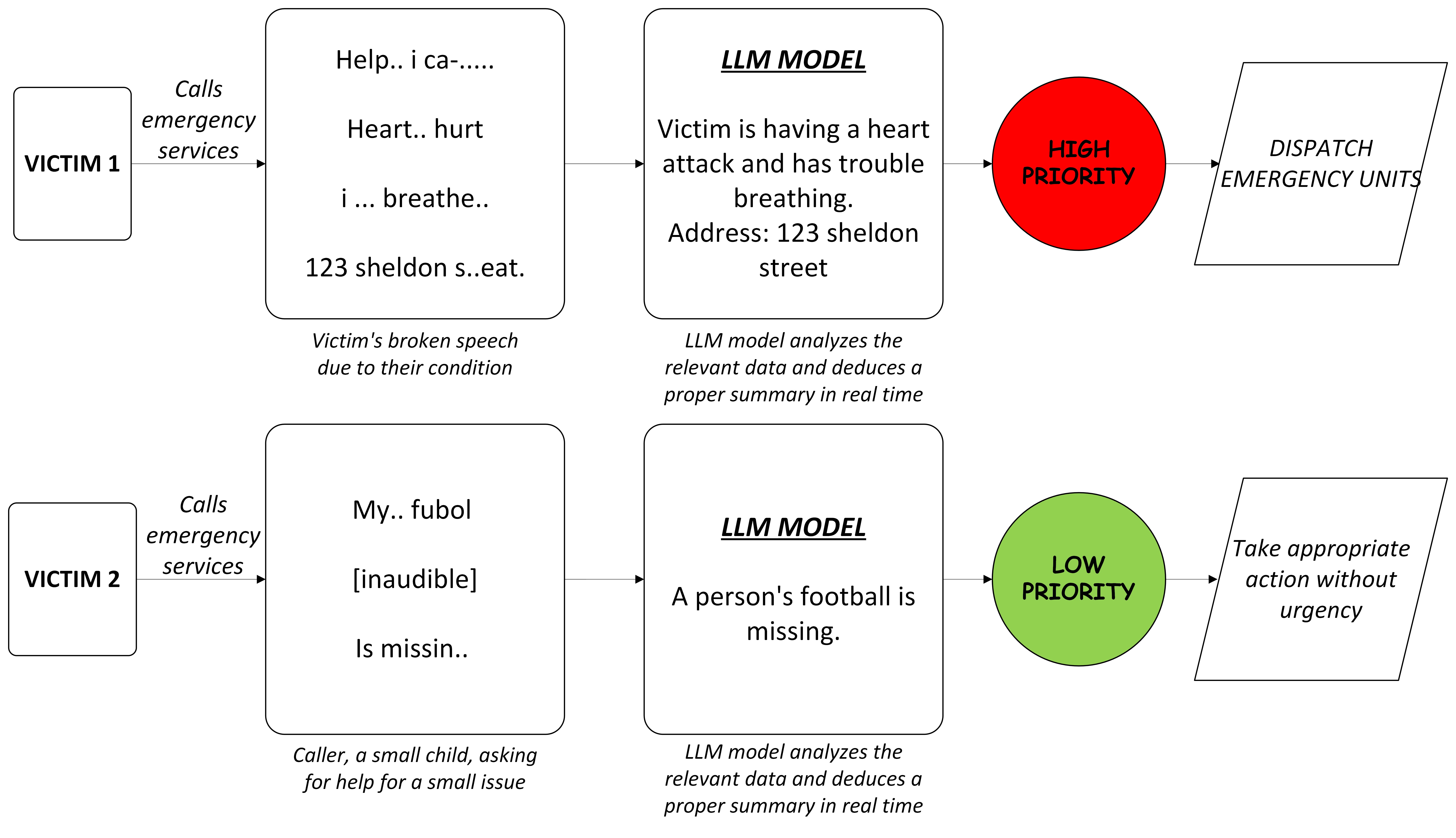}
\caption{LLM based Emergency Call Prioritization System}
\label{fig:roles2}
\end{subfigure}
\caption{Problem statement scenario }
\label{fig:overview}
\end{figure}

\textit{Contextual Reconstruction:} The RAG model processes the caller's potentially broken or unclear language and reconstructs it to form a coherent message, leveraging context from historical emergency data or predefined phrases. This step ensures that the essential details of the emergency are preserved and clarified for accurate interpretation.

\textit{Prioritization of Calls:} The system prioritizes each based on urgency if multiple calls are received simultaneously. This prioritization is determined by analyzing the reconstructed information for key severity indicators, such as specific keywords or emotional cues, enabling the system to rank high to low-urgency cases.

\textit{Dispatch Assignment:} Finally, the prioritized and clarified call information is forwarded to the appropriate dispatcher for action, ensuring that critical cases receive immediate attention.
This study proposes leveraging an LLM to develop an advanced emergency call-handling system. The system can enhance communication clarity, prioritize urgent cases, and streamline response workflows using real-time audio transcription, RAG for enhanced contextual understanding, and rule-based severity classification mechanisms. 

The model aims to reduce response times, improve communication clarity, prioritize urgent cases, and streamline response workflows with actionable information by real-time transcription and contextual understanding with RAG and severity classification. We further describe the model as follows:

\textbf{Intent Prediction}: The LLM-based model predicts the most likely intent \( I_{\text{pred}} \) from a set of possible intents \( I_1, I_2, \ldots, I_k \), based on the transcribed message and contextual information retrieved by the RAG model:
\begin{equation}
I_{\text{pred}} = \arg\max_{i} P(I_i \mid T(X), R)
\end{equation}
where \( P(I_i \mid T(X), R) \) represents the probability of intent \( I_i \) given the transcription \( T(X) \) and retrieved context \( R \).

\textbf{Real-Time Transcription}: Let \( X = \{x_1, x_2, \ldots, x_n\} \) represent a sequence of audio frames from a call. The transcription function \( T \) maps this sequence to text, producing a transcription \( T(X) = \{t_1, t_2, \ldots, t_m\} \), where \( t_i \) represents words or phrases identified in the audio sequence. This transcription serves as the input for the subsequent steps in the system.

\textbf{Contextual Understanding with RAG}: Given the transcription \( T(X) \), the RAG model leverages retrieval-based information \( R \) to fill contextual gaps, producing an enhanced interpretation \( I \) of the emergency situation:
\begin{equation}
I = f_{\text{RAG}}(T(X), R)
\end{equation}
where \( f_{\text{RAG}} \) is a function that combines the transcription \( T(X) \) with relevant retrieved information \( R \) from past emergencies or contextual databases to provide a more complete understanding.

\textbf{Severity Classification}: Using the interpretation \( I \), the system computes a severity score \( S \) based on key features such as keywords \( K \), emotion score \( E \), and context \( C \):
\begin{equation}
S = w_K K + w_E E + w_C C
\end{equation}
where \( w_K \), \( w_E \), and \( w_C \) are weights that reflect the importance of each feature in determining severity. A threshold \( \theta \) is applied to classify the emergency level:
\begin{equation}
\text{Severity Level} = 
\begin{cases} 
   \text{High} & \text{if } S \geq \theta_H \\
   \text{Moderate} & \text{if } \theta_M \leq S < \theta_H \\
   \text{Low} & \text{if } S < \theta_M 
\end{cases}
\end{equation}
where \( \theta_H \) and \( \theta_M \) are predefined severity thresholds.

\textbf{Call Prioritization}: The system assigns a priority score \( P \) to each call based on factors such as severity score \( S \), call frequency \( F \), and caller distress level \( D \):
\begin{equation}
P = w_S S + w_F F + w_D D
\end{equation}
where \( w_S \), \( w_F \), and \( w_D \) are weights that reflect the importance of these factors. The calls are ranked based on the computed priority score \( P \), ensuring that higher priority cases receive immediate attention.

\textbf{VoIP Protocol and Network Modeling:}
In real-time communication systems, VoIP performance is subject to packet loss and bandwidth limitations. Let \( R_p \) be the packet reception rate, defined as:
\begin{equation}
R_p = 1 - P_l
\end{equation}
where \( P_l \) represents the packet loss probability, which can be modeled as a function of random and bursty loss patterns:
\begin{equation}
P_l = P_r + P_b
\end{equation}
where \( P_r \) and \( P_b \) are the probabilities of random and bursty packet loss, respectively. Bandwidth constraints, \( B_{\text{avail}} \), further limit the system's capacity to handle multiple concurrent calls:
\begin{equation}
B_{\text{utilized}} \leq B_{\text{avail}}
\end{equation}
where \( B_{\text{utilized}} \) is the bandwidth required to process audio data for \( N \) active calls:
\begin{equation}
B_{\text{utilized}} = N \times B_{\text{call}}
\end{equation}
where \( B_{\text{call}} \) representing the bandwidth consumed by a single call. If \( B_{\text{utilized}} > B_{\text{avail}} \), call quality degrades, leading to poor signal quality and reduced transcription accuracy.

Fig.~\ref{fig:usecase1} shows the LLM-driven framework for emergency response. It shows the entire cycle of processes that happen once a heart attack victim places an emergency call.
Fig.~\ref{fig:roles2} depicts various scenarios of victims contacting emergency services, with the system prioritizing each case based on urgency.
Fig.~\ref{fig:overview} is an overview of the emergency response system.
This study introduces an innovative system to address these challenges by combining three key components: real-time transcription, intent prediction, and rule-based severity classification. \textcolor{black}{The system uses an advanced LLM to process the transcribed conversation and predict the caller’s intended message}, even when the information is incomplete or unclear. This capability is precious when the caller is distressed, disoriented, or unable to fully describe the emergency's nature.
To improve decision-making and prioritize emergency responses, the system includes a severity classification mechanism that assesses the predicted transcript against predefined rules. These rules are based on keywords and contextual cues that indicate the seriousness of the situation. 

For example, keywords such as "heart attack" or "fire" would trigger a high-severity classification, prompting immediate action from first responders.
The proposed system provides a comprehensive solution for enhancing emergency call management by integrating real-time transcription, predictive modeling, and severity assessment. Its goal is to reduce the time required to assess emergencies, improve the accuracy of severity classifications, and ensure that high-risk situations receive the attention they need as quickly as possible.
This study improves emergency communications systems by showcasing how AI can be utilized in critical real-world scenarios. The system's capability to handle evolving and incomplete information represents a significant advancement over existing approaches, positioning it as a valuable tool for emergency response teams worldwide.

\section{Data Description and Preprocessing}
The emergency call transcripts dataset is created by SpikeCodes and sourced from Hugging Face public repositories \cite{spikecodes_911_call_transcripts}. The dataset includes transcripts of emergency calls featuring conversations between the victim and respondent. This dataset aims to facilitate research in \textcolor{black}{NLP} for predicting what the victim is trying to convey and to reduce the latency between the victim and the respondent. Additionally, the dataset is intended to help predict the criticality of the call.
The dataset contains approximately \texttt{518} detailed conversations. For this project, only the first message from the respondent was extracted and combined with the first two consecutive messages from the victim to form a single response. This approach was chosen to provide a complete response, as the first message alone may not contain sufficient information or context. In cases where the conversation did not have at least one respondent message followed by two victim messages, those conversations were excluded from the analysis to maintain consistency.
Let \( Q \) represent the message from the respondent, and let \( A_1, A_2 \) be the first and second messages from the victim, respectively. The combined response \( R \) is modeled as follows:
\begin{equation}
R = Q \oplus A_1 \oplus A_2
\end{equation}
where \( \oplus \) represents the concatenation operation that joins the messages into a single response. In cases where only one victim message is present, such as \( A_2 = \emptyset \), the conversation is excluded to ensure consistency and completeness in the dataset.
The dataset has been created in our study to explore real-time emergencies, enhance emergency call handling, and classify the criticality of the calls. In the first step, the dataset was loaded from Hugging Face, which provides an interface for handling large datasets in a streamlined manner. The dataset was converted to a DataFrame, denoted as \( D \), which was used for analysis and data manipulation. Each row contains a message from the emergency call respondent as (Q) and is combined with the first two consecutive messages from the victim as \( A_1 \) and \( A_2 \).
The data filtering process is mathematically represented as follows, where \( D' \) represents the cleaned dataset with only complete conversations:
\begin{equation}
\label{Eq_D`}
D' = \{ R_i \in D \mid |R_i| \geq 2 \}
\end{equation}

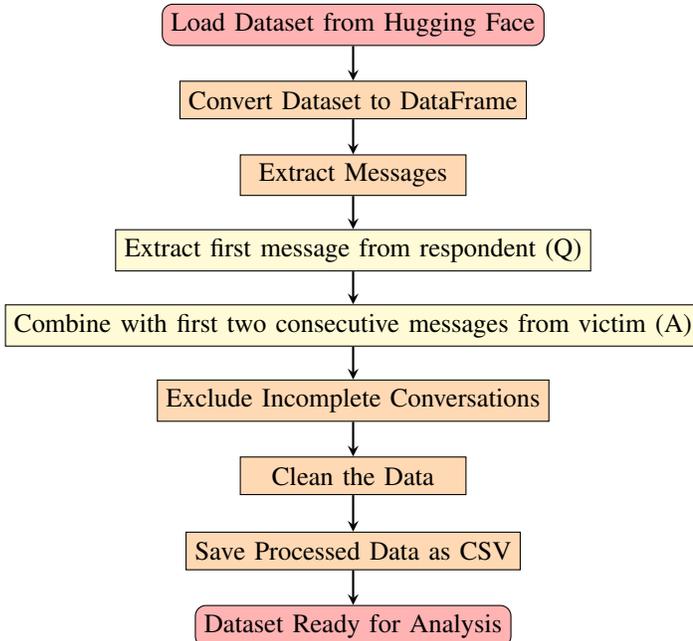
\begin{figure}[h]
\centering
\begin{tikzpicture}[node distance=1cm] % Adjusted node distance for consistency
% Define styles
\tikzstyle{startstop} = [rectangle, rounded corners, minimum width=3cm, minimum height=0.5cm, text centered, draw=black, fill=red!30]
\tikzstyle{process} = [rectangle, minimum width=3cm, minimum height=0.5cm, text centered, draw=black, fill=orange!30]
\tikzstyle{subprocess} = [rectangle, minimum width=2.5cm, minimum height=0.5cm, text centered, draw=black, fill=yellow!20]
\tikzstyle{arrow} = [thick,->,>=stealth]

% Nodes
\node (start) [startstop] {Load Dataset from Hugging Face};
\node (step1) [process, below of=start] {Convert Dataset to DataFrame};
\node (step2) [process, below of=step1] {Extract Messages};
\node (substep1) [subprocess, below of=step2] {Extract first message from respondent (Q)};
\node (substep2) [subprocess, below of=substep1] {Combine with first two consecutive messages from victim (A)};
 \node (step3) [process, below of=substep2] {Exclude Incomplete Conversations};
\node (step4) [process, below of=step3] {Clean the Data};
\node (step5) [process, below of=step4] {Save Processed Data as CSV};
\node (end) [startstop, below of=step5] {Dataset Ready for Analysis};
% Arrows
\draw [arrow] (start) -- (step1);
\draw [arrow] (step1) -- (step2);
\draw [arrow] (step2) -- (substep1);
\draw [arrow] (substep1) -- (substep2);
\draw [arrow] (substep2) -- (step3);
\draw [arrow] (step3) -- (step4);
\draw [arrow] (step4) -- (step5);
\draw [arrow] (step5) -- (end);

\end{tikzpicture}
\caption{Data description and preprocessing steps}
\label{fig:preprocessing_flowchart}
\end{figure}

Note (\ref{Eq_D`}) ensures that only rows with sufficient data, i.e., a respondent message and two victim messages, are included in the final analysis. Data normalization was not required, as the project focused on preserving the natural language form. However, basic dataset cleaning was performed. The final dataset, which consists of question-answer pairs, is saved in CSV format for further analysis \cite{Venkateshperumal2024RAG}.
If only one victim message was available, it might not contain sufficient information about the emergency. In cases where the conversation did not include at least one respondent message followed by two victim messages, those conversations were excluded from the analysis to maintain consistency.
Data normalization was not required, as the project focused on preserving the natural language form. However, basic dataset cleaning was performed. The final dataset, which consists of question-answer pairs, is saved in CSV format for further analysis \cite{Venkateshperumal2024RAG}.
Fig.~\ref{fig:preprocessing_flowchart} outlines the data preprocessing workflow. Table~\ref{tab:sample of emergency call} shows various samples of how the emergency call transcriptions look like.

\begin{table}[ht]
\centering
\caption{Sample of Emergency Call}
\label{tab:sample of emergency call}
\begin{tabular}{|c|p{4cm}|}
\hline
\textbf{Q} & \textbf{A} \\
\hline
9-1-1, what's your emergency? & I'm at West High School. There's a guy with a gun. West High. \\
\hline
9-1-1, what's your emergency? & Hi, you need to get the police. Just ran over and they said that the young boy just shot his mom. \\
\hline
\hline
9-1-1, what's your emergency? & It's 188 and 92 Indian Springs. I just shot my mother and my nephew. \\
\hline
9-1-1, what's your emergency? & He's breaking into my house! I don't know who he is! \\
\hline
\end{tabular}
\end{table}

\begin{figure*}[]
\centering
\includegraphics[width=6.50in]{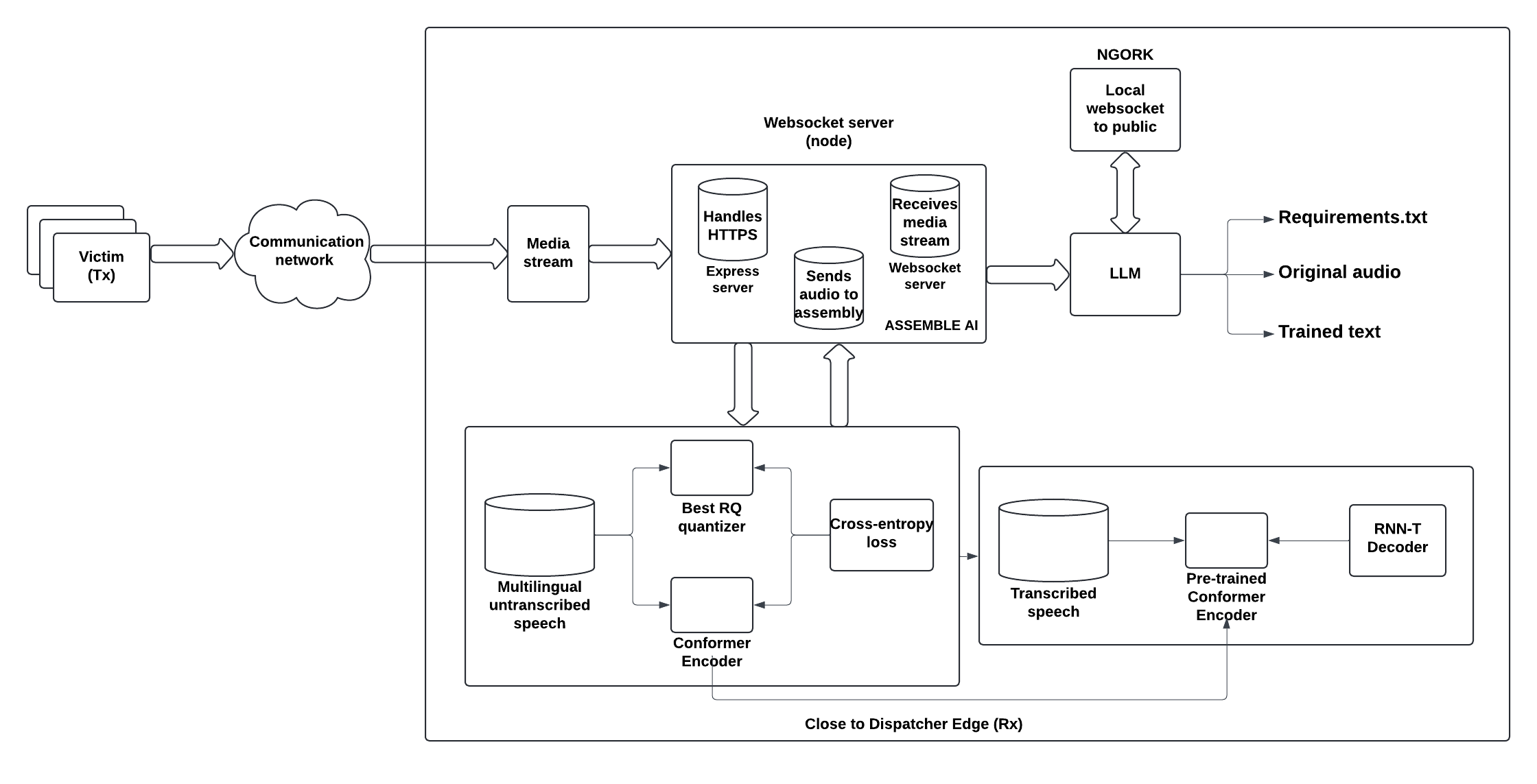}
\caption{System Design}
\label{fig:system design}
\end{figure*}

\section{Proposed Methodology}
In this system, we discuss the proposed solution approach of the LLM-based emergency service model. As illustrated in Fig.~\ref{fig:system design}, the system design depicts an intricate flow of interconnected functionalities working in tandem within this system. It begins with a victim transmitting an audio plea via communication networks packet loss, muffled voice, and poor signal quality, then processed through a WebSocket server and sent through a complex LLM pipeline. Key components breaking down the audio include real-time speech transcription,  audio parsing, and signal quantization, leveraging cutting-edge neural models such as Conformer Encoders and Recurrent Neural Network Transducer decoders. These advanced techniques diligently convert packet loss, muffled voice, and poor signal quality speech into understandable text and discern contextual meaning with notable accuracy. Ultimately, this allows distressed individuals to secure urgent aid when needing assistance most.

\subsection{Real-Time Audio Transcription with \texttt{Twilio} and \texttt{AssemblyAI}}
\subsubsection{\texttt{Twilio} Media Stream for Real-Time Audio Capture}
\texttt{Twilio}’s Media Stream \texttt{API} is the core component that captures audio from a phone call in real time. This media stream is necessary to convert live speech into a format suitable for transcription. A \texttt{Twilio} phone number is required to initiate the process. When receiving an incoming call, \texttt{Twilio} triggers a webhook, directing the call to a specific server endpoint. In this setup, the endpoint is a \texttt{WebSocket} server capable of handling live audio streams. \texttt{Twilio} streams the audio as \texttt{base64-encoded} pulse code modulation (PCM).
The \texttt{Twilio} phone number is configured to respond with TwiML, which directs the audio stream, including any instances of packet loss, muffled voice, or poor signal quality, to the \texttt{WebSocket} server. This is accomplished through a simple POST request that instructs \texttt{Twilio} on how to handle the media stream. Once the call is connected, \texttt{Twilio} begins streaming audio over \texttt{WebSocket}, ensuring a continuous real-time connection where the voice data is received in small chunks as \texttt{base64-encoded} strings.

\subsubsection{\texttt{WebSocket} Server Setup}
The \texttt{WebSocket} server is built using \texttt{Node.js} to handle incoming audio streams from \texttt{Twilio}. It listens for messages from \texttt{Twilio} and decodes the audio data, which is sent in base64 format, into a buffer. This buffer is then forwarded to the transcription service. \texttt{Twilio}'s media stream sends four types of events: \textit{connected}, \textit{start}, \textit{media}, and \textit{stop}. The \textit{media} event carries the actual audio data transcribed in real-time. When a \textit{media} event is received, the server extracts and decodes the audio chunk, sending it to \texttt{AssemblyAI} for transcription. This step ensures the audio is processed continuously and quickly, almost in real-time.

\subsubsection{Real-Time Transcription with \texttt{AssemblyAI}}
\texttt{AssemblyAI}'s speech-to-text \texttt{API} powers the transcription process. After receiving the  poor quality audio from \texttt{Twilio}, \texttt{AssemblyAI} converts it into text transcripts in real-time.

\paragraph{Connecting to \texttt{AssemblyAI}}
An \texttt{API} key is needed to authenticate requests to \texttt{AssemblyAI}. The connection is established using the \texttt{RealtimeTranscriber} from \texttt{AssemblyAI}’s \texttt{API}, with the audio encoding and sample rate configured to match \texttt{Twilio}’s specifications.

\paragraph{Handling Transcripts}
\texttt{AssemblyAI} generates two types of transcripts: partial and final. Partial transcripts are produced continuously as the audio is processed, while final transcripts are generated when the speaker finishes talking. The system uses only the final transcripts for further processing, sending them to the language (RAG) model for analysis. Throughout the call, the server transcribes audio in real-time, line-by-line, and keeps the connection open until the media stream ends. The connection remains open throughout the phone call, continuously transcribing audio until the media stream stops.

\subsection{Response Generation using RAG}
Once the transcript is complete, the system uses a RAG model to generate predictive responses that are contextually relevant to the conversation.

\subsubsection{RAG Model Architecture}
The RAG model leverages a combination of two powerful components:
\begin{itemize}
    \item \textbf{Retriever:} This part of the model searches a knowledge base (such as indexed documents) to find information that matches the conversation’s context.
    \item \textbf{Generator:} Using the BART (Bidirectional and Auto-Regressive Transformers) architecture, the generator creates responses based on the current conversation and the retrieved documents.
\end{itemize}

\subsubsection{RAG Model Setup}
The RAG model is fine-tuned explicitly for emergency call transcripts in this system. It uses a custom FAISS index, built from a database of emergency calls. When a new transcript is processed, the RAG model searches this index to find similar situations and uses those results to help generate the response.
\begin{verbatim}
def load_csv_data(file_path):
    data = pd.read_csv(file_path)
    data['combined_text'] = data.apply(
    lambda row: ' '.join(row.values.astype
    (str)), axis=1)
    return data

# Create TF-IDF embeddings
def create_embeddings(data):
    vectorizer = TfidfVectorizer()
    embeddings = vectorizer.fit_transform
    (data['combined_text']).toarray()
    return embeddings, vectorizer

# Build FAISS index
def build_faiss_index(embeddings):
    dim = embeddings.shape[1]
    index = faiss.IndexFlatL2(dim)
    index.add(embeddings)
    return index
\end{verbatim}

\subsubsection{Fine-Tuning RAG for Emergency Transcripts}
We fine-tune the RAG model using a dataset of real emergency call transcripts to make it suitable for emergencies. This dataset includes pairs of typical questions from operators and responses from callers, helping the model learn to respond appropriately in these scenarios.

\subsubsection{Response Generation Process}
Once the RAG model is fine-tuned, it generates responses based on new transcripts. When the system receives a final transcript, the RAG model processes it and generates a predicted continuation of the conversation.

\begin{verbatim}
def retrieve_entries(query, index,
data,vectorizer, k=5):
    query_vec = vectorizer.transform([query])
    .toarray()
    _, indices = index.search(query_vec, k)
    retrieved_texts = data.iloc[indices[0].
    astype(int)]['combined_text'].tolist()
    return retrieved_texts

# Query the GPT-3 model with the retrieved 
context
def chat_gpt(query, retrieved_texts):
    context = "\n".join(retrieved_texts)
    prompt = f"Context:\n{context}\n\n
    Given the partial transcript: '{query}', 
    
    predict what the speaker is most 
    likely saying."
    
    response = openai.ChatCompletion.create(
        model="gpt-3.5-turbo",
        messages=[{"role": "user", "content":
        prompt}],
        max_tokens=150,
        temperature=0.7
    )
    return response.choices[0]['message']
    ['content'].strip()
    \end{verbatim}

\section{Severity Classification}
Accurately determining a situation's urgency in emergency communications systems is essential for efficiently deploying resources. This module analyzes transcripts from emergency phone calls and assigns a severity level based on the conversation’s content. By categorizing the severity, the system aids in prioritizing emergencies, ensuring that high-risk situations receive immediate attention while less urgent cases are handled accordingly.

\subsection{Rule-Based Severity Classification}
The system utilizes a rule-based method to determine the seriousness of emergencies by analyzing the transcribed conversations. This approach is simple yet effective for identifying critical events commonly linked to severe situations, such as fires, shootings, or accidents.

\subsubsection{Keyword Detection}
The rule-based classifier scans the transcript for specific keywords that indicate urgency. Two sets of keywords are employed:
\begin{itemize}
    \item \textbf{Severe Keywords:} These words are typically associated with life-threatening or urgent scenarios. If these words appear in the transcript, the system automatically flags the emergency as high priority (severity level 4). Common examples include: \textit{gun, stabbing, shooting, fire, accident, emergency, death, killed}.
    \item \textbf{Mild Keywords:} These words relate to less pressing situations and are not likely to pose an immediate risk to life or property. The emergency is classified as low priority (severity level 1) if these words are present. Examples include: \textit{noise, neighbor, pet, minor}.
    \item \textbf{Moderate Severity:} If no severe or mild keywords are detected, the system defaults to assigning a moderate priority level (severity level 2).
\end{itemize}
\begin{algorithm}
\caption{Real-Time Emergency Communication System with RAG and Severity Classification}
\textbf{Input:} Real-time audio stream from an emergency call \\
\textbf{Output:} Predicted response using RAG and severity classification
\begin{algorithmic}[1]
\State \textbf{Step 1: Initialize System Components}
    \State Load and configure the RAG model with a pre-trained tokenizer, retriever, and generation model.
    \State Load the severity classification model and define keyword lists 
    \State Set up the  \texttt{WebSocket} server to receive audio streams.
    \State Initialize the \texttt{AssemblyAI} transcriber for real-time transcription.
\State \textbf{Step 2: \texttt{WebSocket} Connection \texttt{(\texttt{Node.js})}}
    \State On \texttt{WebSocket} connection:.
        \State Log connection information (e.g., Session ID). On receiving 'start' event:
        \State Log " media stream started." On receiving 'media' event:
        \State Convert the audio payload into a suitable format.
        \State Send audio data to \texttt{AssemblyAI} transcriber for real-time transcription.        
\State \textbf{Step 3: Real-Time Transcription \texttt{(\texttt{Node.js} \& \texttt{AssemblyAI})}}
        \State Send the final transcript to the \texttt{Flask} \texttt{API} using \texttt{HTTP POST} request.
\State \textbf{Step 4: Process Transcript \texttt{(\texttt{Flask})}}
    \State Receive the transcript from the \texttt{HTTP POST} request.
    \State Generate Predicted Response using RAG:
        \State Tokenize the transcript using RAG tokenizer.
        \State Generate a contextually enriched response using the RAG model.
\State \textbf{Step 5: Return Results to Client \texttt{(\texttt{Flask})}}
    \State Package the predicted response and severity level into a \texttt{JSON} object.
    \State Send the \texttt{JSON} object back to the \texttt{Node.js} server.
\State \textbf{Step 6: Display Results and Handle Response \texttt{(\texttt{Node.js})}}
    \State Receive the \texttt{JSON} response from the \texttt{Flask} \texttt{API}.
    \State Log the RAG-generated response and severity level.
    \State Output the results.
\end{algorithmic}
\end{algorithm}

\subsection{Classification Process}
To prevent issues with capitalization, the system first converts the entire transcript to lowercase. It then begins by scanning for any severe keywords. If a severe keyword is found, the system immediately assigns a severity score of 4, signaling a critical emergency. If no severe keywords are present, the system checks for mild keywords. The system assigns a severity score of 1 if any mild keyword indicates a lower-priority emergency.
While this approach works well, it does rely heavily on exact keyword matches. This can sometimes lead to misclassification if the transcript uses a synonym or related word that isn’t in the keyword list (e.g., "blaze" instead of "fire").
% Begin the table
\begin{table}[htbp]
\centering
\caption{Partial Transcripts and LLM Predictions}
\begin{tabular}{|p{3cm}|p{5cm}|}
\hline
\textbf{Partial Transcript} & \textbf{LLM Prediction} \\
\hline
acid attack face burning help me & The speaker is most likely reporting an acid attack where their face is burning and they need help. \\
\hline
Bleeding, severe pain. Help me. Help. & The speaker is most likely describing a situation where they are experiencing severe bleeding and pain, and are in urgent need of assistance. \\
\hline
smoke building on fire & The speaker is most likely reporting a fire in a building with smoke coming from it. \\
\hline
it's neighbor disturbing nice neighbor disturbing what can he do & "It's my neighbor. He's disturbing. He's a nice neighbor, but he's disturbing. What can he do?" \\
\hline
cat and away were not able to find cat & The speaker is most likely saying "My cat ran away and we are not able to find it." \\
\hline
biocycle last someone's old bicycle after the last bicycle & Based on the partial transcript provided, it is difficult to accurately predict what the speaker is most likely saying. The words "Biocycle" and "bicycle" do not seem to have a clear connection in this context. It is possible that the speaker is talking about a stolen bicycle or some kind of cycling event, but without more information it is hard to determine the exact meaning. \\
\hline
i've complained dog barking too much noise complained dog barking bye bye & The speaker is most likely reporting a noise complaint about a barking dog. \\
\hline
covered her gun short school gun someone is having gun in school having gun & The most likely statement the speaker is trying to convey is "There is someone with a gun at the school, I heard gunshots." \\
\hline
\hline
Somebody into my house & 9-1-1, what's your emergency? Somebody just came into my house. Help me. I'm afraid. \\
\hline
i'm on west street someone took my son afraid they kidnapped suffrage kid & I'm on West street. Someone took my son. I'm afraid they kidnapped him. \\
\hline
she middle divorce i am sure she supposedly went to morning run no one heard she's missing i'm afraid & The speaker is most likely reporting that their friend, who is in the middle of a divorce, has gone missing after going for a morning run and no one has heard from her. They are expressing fear and concern for their friend's safety. \\
\hline
need ambulance guys stayed with me last night he's not responding & The speaker is most likely reporting that someone they were with the previous night is unresponsive and in need of medical attention. \\
\hline
i don't know some lady random lady sidewalk bleeding on the street & The speaker is most likely reporting a situation where they have come across a random lady lying on the sidewalk and bleeding on the street. \\
\hline
back pain help ambulance not moving & The speaker is most likely saying "I'm feeling pain in my back." and is requesting an ambulance for a medical emergency. The person they are calling about is not moving. \\
\hline
and my mama fell down help & The speaker is most likely reporting that their mother has fallen and is in need of help. \\
\hline
My neighbor. I think someone is killing her with the gun. She ran to my place. I'm afraid. There is gunshot. I need my lawyer. That lawyer. Help me. & The speaker is most likely reporting a situation where their neighbor is in danger and believes someone is trying to harm her with a gun. The neighbor has sought refuge at the speaker's place, and the speaker is afraid for both their neighbor's safety and their own. \\
\hline
i am kidnapped helping i'm on the car they keep left me i am on the road healthy trees i'm afraid panicking car kidnapping help me i am drugged & The speaker is most likely saying: "I have been kidnapped. Please help me. I am in a car. They are not letting me go. I am on the road surrounded by trees. I'm afraid." \\
\hline
That is a dog. The dog is hurt. Dog able not to help the dog. & The speaker is most likely saying that a dog is injured and in need of help, but they cannot assist the dog themselves. \\
\hline
\end{tabular}
\label{tab:transcript_llm_prediction}
\end{table}

\subsubsection{Emotion Classifier Using DistilBERT}
The emotion classifier is built using DistilBERT, a smaller and faster variant of the BERT (Bidirectional Encoder Representations from Transformers) model. DistilBERT retains 97\% of BERT's language understanding capabilities but is much more efficient, making it suitable for real-time applications like emergency call analysis.

The classifier is accessed via the Hugging Face \texttt{transformers} library using the \texttt{pipeline()} function, simplifying the text classification process. A pre-trained model, \texttt{"distilbert-base-uncased-emotion"}, is used for emotion detection.

\begin{itemize}
    \item \textbf{Severe Emotion (Level 4):} If the emotion is classified as sadness or anger, the system considers the situation severe, as these emotions often indicate distress or danger.
    \item \textbf{Mild Emotion (Level 1):} If the emotion is classified as joy, the system considers the situation mild, likely not an emergency.
    \item \textbf{Moderate Emotion (Level 2):} If no severe or mild emotions are detected, the system defaults to a moderate severity level, assuming that the situation may require attention but is not immediately life-threatening.
\end{itemize}

\section{Results and Anslysis}
The system's evaluation results demonstrate its effectiveness in accurately predicting the intent of emergency communications, prioritizing cases based on severity, and generating relevant information using LLM. The system underwent testing across various emergency scenarios, including acid attacks, fires, noise complaints, and gunshot reports. Three primary metrics were employed for the evaluation: Bilingual Evaluation Understudy (BLEU) score, Recall-Oriented Understudy for Gisting Evaluation (ROUGE) score, and Conceptual Precision\footnote{The code of paper is available at \cite{Venkateshperumal2024RAG}}.

% Begin the table
\begin{table}[htbp]
\centering
\caption{BLEU, ROUGE, and Conceptual Precision Scores for Test Cases}
\begin{tabular}{|p{1.5cm}|p{1.5cm}|p{2.5cm}|p{1.5cm}|}
\hline
\textbf{Test Case} & \textbf{BLEU Score} & \textbf{ROUGE Score} & \textbf{Conceptual Precision (\%)} \\
\hline
Heart Attack & 0.0109 & precision=0.2857\newline precision=0.2857 & 100 \\
\hline
Acid Attack & 0.0083 & precision=0.1304\newline precision=0.0870 & 100 \\
\hline
LegBroken & 0.0083 & precision=0.1304\newline precision=0.0870 & 100 \\
\hline
Smoke & 0.0123 & precision=0.1875\newline precision=0.0625 & 100 \\
\hline
Noise Neighbour & 0.1619 & precision=0.55\newline precision=0.55 & 100 \\
\hline
Cat Ran Away & 0.0607 & precision=0.3889\newline precision=0.3333 & 100 \\
\hline
Lost Bicycle & 0.0033 & precision=0.0615\newline precision=0.0615 & 100 \\
\hline
Dog Barking & 0.0000 & precision=0.2308\newline precision=0.1538 & 100 \\
\hline
Gun Shot & 0.0109 & precision=0.2609\newline precision=0.1739 & 100 \\
\hline
Somebody into my house & 0.4471 & precision=0.5\newline recall=1.0\newline fmeasure=0.6667 & 100 \\
\hline
I'm on West street. Someone took my son. & 0.6115 & precision=0.8\newline recall=0.8571\newline fmeasure=0.8276 & 100 \\
\hline
She went missing after a morning run & 0.0061 & precision=0.2045\newline recall=0.4286\newline fmeasure=0.2769 & 100 \\
\hline
Need ambulance, he's not responding & 0.0360 & precision=0.5385\newline recall=0.7778\newline fmeasure=0.6364 & 100 \\
\hline
Random lady bleeding on the street & 0.1044 & precision=0.28\newline recall=0.5385\newline fmeasure=0.3684 & 100 \\
\hline
Back pain, not moving & 0.0123 & precision=0.125\newline recall=1.0\newline fmeasure=0.2222 & 100 \\
\hline
Mother fell down, needs help & 0.01 & precision=0.1176\newline recall=0.2857\newline fmeasure=0.1667 & 100 \\
\hline
Neighbor in danger with gunshot & 0.4005 & precision=0.6111\newline recall=1.0\newline fmeasure=0.7586 & 100 \\
\hline
Kidnapped, in car, drugged & 0.0486 & precision=0.3333\newline recall=0.875\newline fmeasure=0.4828 & 100 \\
\hline
Dog hurt, unable to help & 0.0219 & precision=0.1111\newline recall=1.0\newline fmeasure=0.2 & 100 \\
\hline
\end{tabular}
\label{tab:bleu_rouge_precision}
\end{table}

\begin{figure}[htbp]
\centering
\includegraphics[width=\columnwidth]{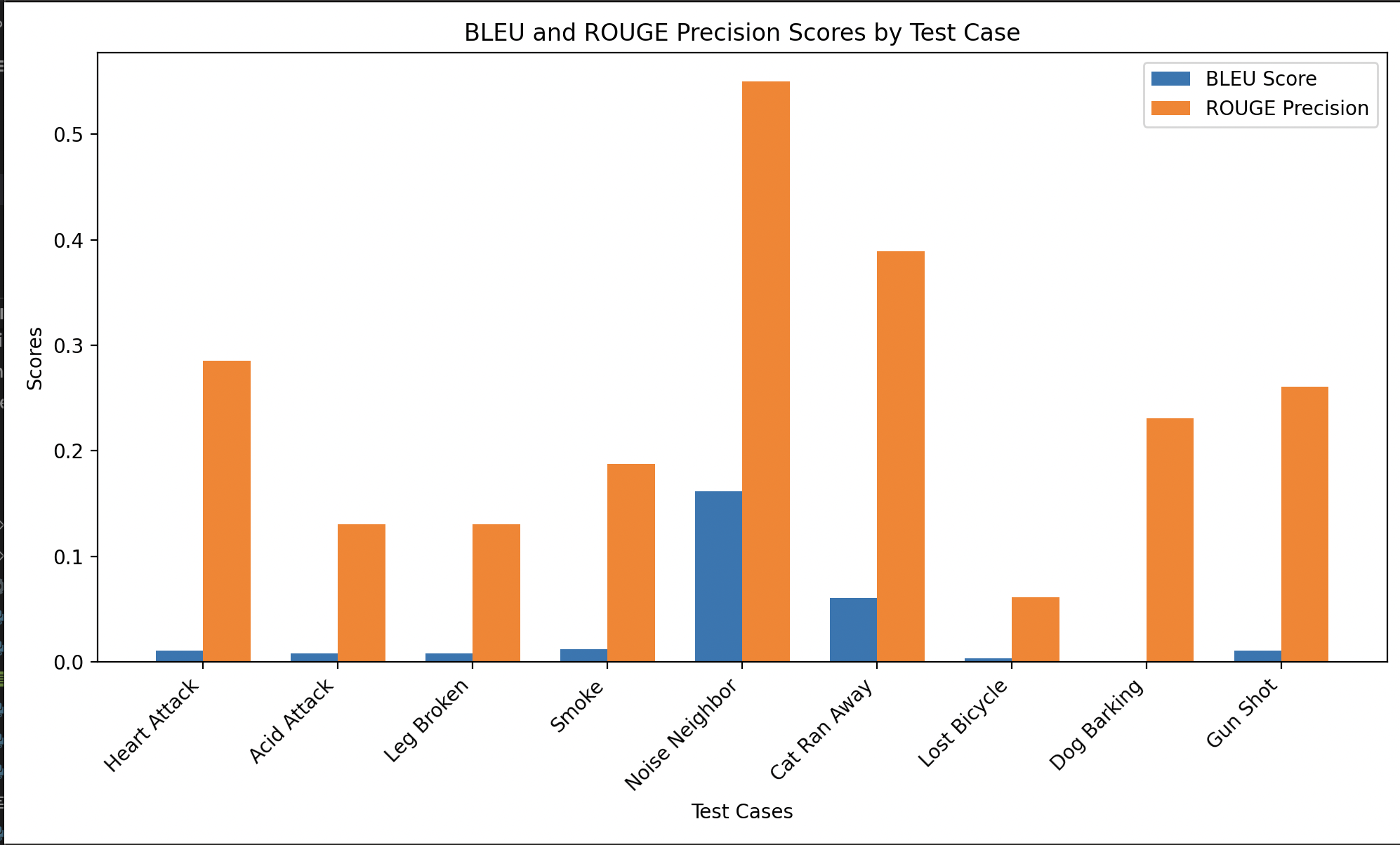} % Adjust the width to \columnwidth
\caption{Scores from various emergency scenarios }
\label{fig:Scores}
\end{figure}

\textit{1. BLEU Score:}
The BLEU score assesses how closely the LLM-generated prediction aligns with the ground truth. In this context, a lower BLEU score is more desirable, as the system aims to produce conceptually relevant and contextually enriched responses rather than exact reproductions of the transcript. The following test cases exemplify this behavior:
Heart Attack (BLEU = 0.0109): The system provided a more detailed response, outlining the symptoms and urgency associated with a heart attack.
Acid Attack (BLEU = 0.0083): While the generated prediction did not match the ground truth word for word, it effectively conveyed the severity and critical elements of the situation, resulting in a low, desirable BLEU score.
Gun Shot (BLEU = 0.0109): The LLM enhanced the description by specifying the sounds of gunshots, adding more relevant details than the transcript alone, thus yielding a low BLEU score.
In these instances, a low BLEU score indicates the model's capability to surpass basic transcription by capturing contextually relevant details and enhancing the system's effectiveness in conveying the situation.
Fig.~\ref{fig:Scores} shows the various BLEU and ROGUE precision scores obtained for test cases involving various emergency scenarios. The system’s performance indicates its ability to infer missing context, particularly in real-time scenarios where packet loss and audio quality may impair exact phrasing. The model's resilience against minor discrepancies in transcription quality, as affected by network conditions, supports its reliability in practical VoIP environments.

\textit{2. ROUGE Score:}
The ROUGE  score measures the overlap of n-grams between the generated output and the ground truth, focusing on recall and f-measure. The results demonstrate that while ROUGE scores vary across different test cases, the system consistently captures essential elements of the transcript:
Heart Attack (ROUGE precision = 0.2857): The system accurately included core concepts such as "heart attack" and symptoms like "pain on the left side."
Smoke (ROUGE precision = 0.1875): The system captured keywords like "fire" and "smoke," although the lower precision indicates that the generated prediction contained additional context not present in the original transcript.
Noise Neighbor (ROUGE precision = 0.55): In this case, the system achieved higher ROUGE precision, capturing most of the critical details about the noise complaint, a relatively straightforward situation. This robustness is essential in cases where audio quality is compromised due to bandwidth limitations or packet loss, as the model can still capture critical information even if exact n-gram matches are affected.

\textit{3. Conceptual Precision:}
The Conceptual Precision metric evaluates how well the system captures the intent and fundamental concepts of the situation, regardless of exact phrasing. In all test cases, the system achieved 100\% Conceptual Precision, indicating that the LLM successfully interprets the key aspects of each emergency and delivers contextually relevant outputs.
In the Acid Attack scenario, the model correctly recognized the severity of the situation and provided a detailed prediction aligned with the urgency of the request for help.
In the Gun Shot case, the model went beyond the raw transcript to infer and predict that gunshots were heard, adding valuable context that might be overlooked.

\begin{figure}[htbp]
\centering
\includegraphics[width=\columnwidth]{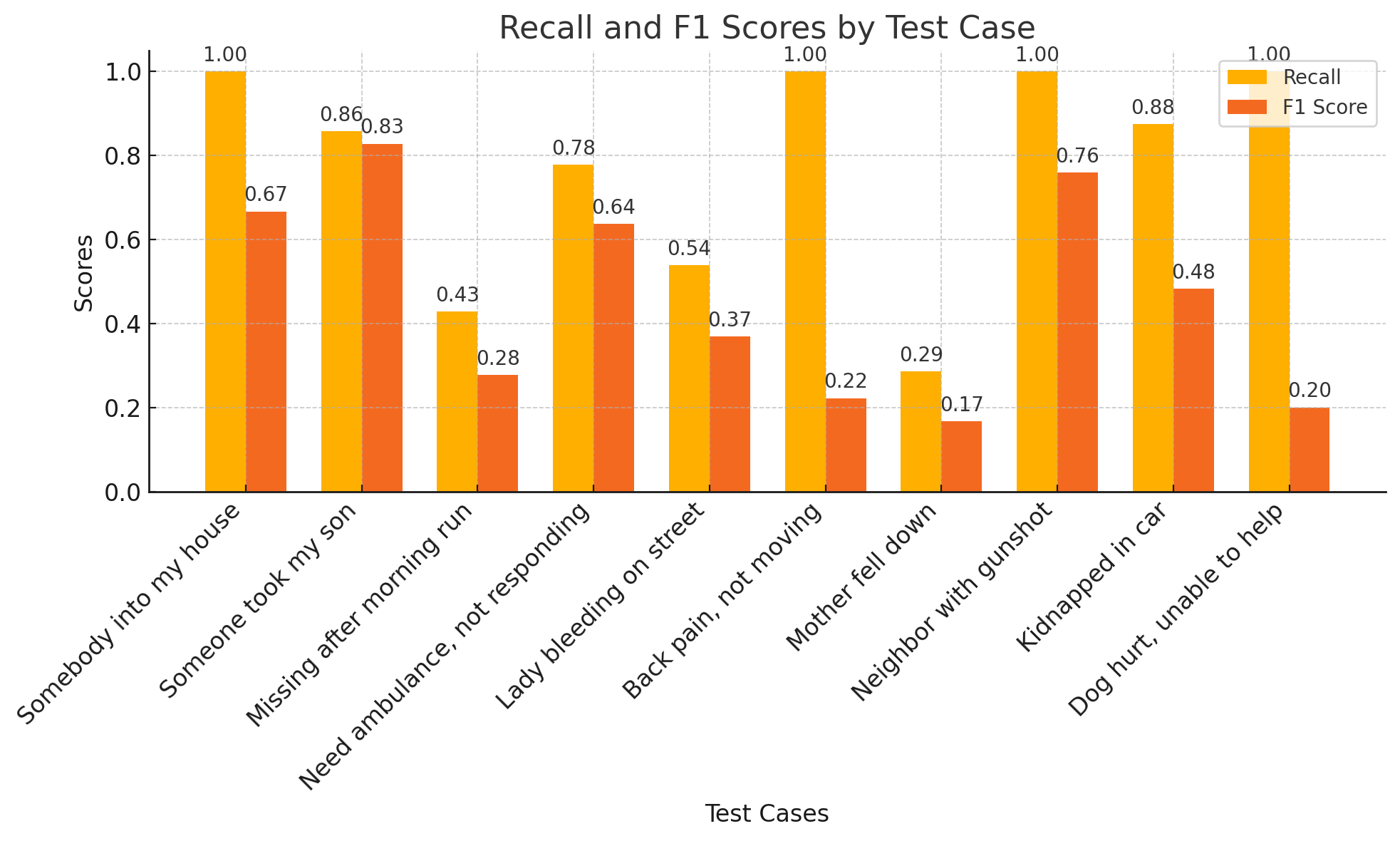} % Adjust the width to \columnwidth
\caption{Comparison of Recall and F1 Scores  }
\label{fig:Comparison.}
\end{figure}

\textit{4. F1 Score and Recall:}
In test cases such as "Someone in my home," "Back pain, cannot move," "Neighbor with a gunshot wound," and "Dog injured, unable to assist," the system achieved recall scores nearing 1.0. This impressive recall signifies that the model successfully identifies nearly all pertinent information. 
In scenarios like "Missing after morning jog" and "Mother fell, requires assistance," the system produced lower F1 scores. This highlight instances where minor transcription errors, potentially due to VoIP limitations, impacted the completeness of the information captured. These cases suggest that while the model is effective in high-stress emergencies, additional refinement is needed to handle situations where packet loss or bandwidth constraints obscure parts of the audio.
Across all test cases, the system attained a Conceptual Precision of 100\%. This metric evaluates the system’s capability to grasp the overall intent or meaning of the emergency transcript rather than focusing on the exact wording.
The analysis of recall and F1 scores indicates that the system excels at extracting critical information while maintaining high conceptual precision, particularly in urgent emergencies.
Fig.~\ref{fig:Comparison.} depicts the comparison of Recall and F1 scores across various emergency scenarios.

\textit{5 Severity Classification:}
Among the eight actual Severe cases, the model successfully identified 5 cases as Severe (62.5\% of Severe cases), showing that the model generally effectively recognizes high-urgency scenarios.
Nevertheless, 3 Severe cases (37.5\%) were incorrectly categorized as Moderate. This misclassification indicates that while the model proficiently recognizes urgent situations, it sometimes underestimates the severity, mainly when severe indicators are less apparent.
The model did not categorize any cases as Mild, wrongly classifying both Mild cases as Moderate. The system’s performance in real-time environments, particularly under packet loss or low bandwidth conditions, reveals that while it is sensitive to critical keywords, it may underclassify severity if audio degradation masks these cues. By refining the classifier to handle subtle or fragmented inputs better, we can potentially reduce these misclassifications.
The dataset lacked any confirmed cases at the Moderate severity level, hampers assessing the model's performance regarding moderate cases.
Let the severity classification function \( S \) be defined as:
\[
S = f(K, E, C)
\]
where \( K \) is a vector representing the presence of severity-related keywords, such as "fire," "gunshot," or "injury."
  \( E \) represents an emotion score derived from the caller’s tone or word choices, indicating fear, panic, or urgency levels.
 \( C \) is a contextual factor derived from prior knowledge or historical data, providing additional information about the emergency.
If we introduce weights to each component, we can represent the severity function as a weighted sum:
\[
S = w_K K + w_E E + w_C C
\]
where \( w_K \), \( w_E \), and \( w_C \) are weights that determine the relative importance of each component (keywords, emotion, and context) in the severity classification.
The probability of each severity level \( S = s \) given input features \( X \) is defined as:
\[
P(S = s \mid X) = \frac{\exp(\theta_s X)}{\sum_{j=1}^m \exp(\theta_j X)}
\]
where \( S \) is the severity level, with possible values such as mild, moderate, and severe.
 \( X \) is the feature vector representing the input data (e.g., keywords, emotion score, context).
 \( \theta_s \) is the weight vector associated with severity level \( s \).
 \( m \) is the total number of possible severity levels.
The misclassification penalty function \( \text{Penalty} \) is defined as:
\[
\text{Penalty} = \sum_{i=1}^N \delta_i L(C_i, P_i)
\]
where  \( N \) is the total number of cases.
 \( \delta_i \) is a severity coefficient for case \( i \), which increases for higher-severity misclassifications.
 \( L(C_i, P_i) \) is a loss function, such as cross-entropy loss, that penalizes the difference between the true class \( C_i \) and the predicted class \( P_i \).

The confusion matrix illustrates the model's strengths and weaknesses. The high true positive rate for Severe cases indicates that the model is dependable in recognizing critical situations, which is vital in emergency settings.
Let the confusion matrix \( \mathbf{C} \) be defined as:
\[
\mathbf{C} = \begin{bmatrix} 
    A & B \\ 
    C & D 
\end{bmatrix}
\]
where \( A \) is the number of severe cases correctly classified as severe.
 \( B \) is the number of severe cases incorrectly classified as mild or moderate.
 \( C \) is the number of mild or moderate cases incorrectly classified as severe.
 \( D \) is the number of mild or moderate cases correctly classified.
Using the values from the confusion matrix, we can calculate key performance metrics as follows:
Precision,  \( (P) \) is expressed as follows:
\[
P = \frac{A}{A+ C}
\]
Recall, \( R \) is expressed as follows:
\[
R = \frac{A}{A + B}
\]
\( F_1 \) is expressed as follows:
\[
F_1 = 2 \times \frac{P \times R}{P + R}
\]

\begin{figure}[htbp]
\centering
\includegraphics[width=\columnwidth]{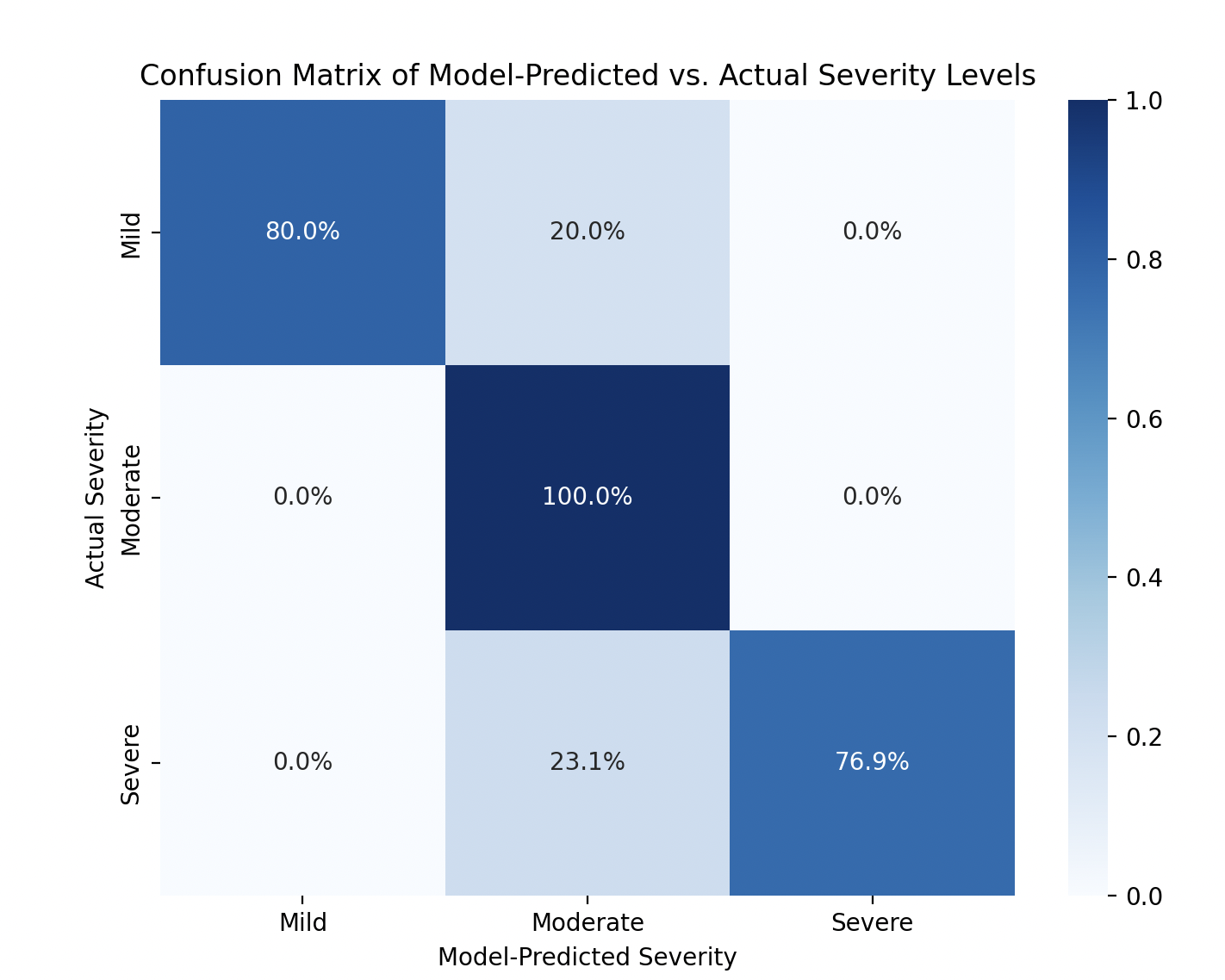} % Adjust the width to \columnwidth
\caption{Confusion Matrix of Model-Predicted vs. Actual Severity Levels in Emergency Scenario }
\label{fig:confused}
\end{figure}

The confusion matrix in Fig.~\ref{fig:confused} the model's performance across three severity levels: Mild, Moderate, and Severe. For Mild severity (Row 1), the model correctly classified 80\% of cases as Mild, while 20\% were misclassified as Moderate, with no cases misclassified as Severe. For Moderate severity (Row 2), the model achieved 100\% correct classification, accurately identifying all Moderate cases, with no instances misclassified as either Mild or Severe. For Severe severity (Row 3), the model correctly predicted 76.9\% of cases as Severe but misclassified 23.1\% as Moderate, with no instances misclassified as Mild.

This analysis reveals the model’s strengths in correctly identifying Severe cases and avoiding false positives for Mild severity, as it never misclassifies a non-Mild case as Mild. However, there is still some underestimation of severity, as a portion of Severe cases were misclassified as Moderate. The model demonstrates strong performance with Moderate cases, correctly classifying all instances.
% Table~\ref{tab:bleu_rouge_precision} presents BLEU, ROUGE, and Conceptual Precision scores across various emergency scenarios, evaluating the model’s accuracy in generating contextually relevant responses.
% Table~\ref{tab:summary_values} summarizes critical performance metrics, including BLEU, ROUGE Precision, Conceptual Precision, and Misclassification Rate, for selected emergency test cases to assess model efficacy in prioritizing responses.
Upon reviewing misclassified cases, the model occasionally underestimates severity in scenarios with subtle or indirect language cues, particularly when cases lack precise keywords associated with higher severity (e.g., ‘disturbance’ instead of ‘gunshot’). The model’s sensitivity to indirect expressions may also lead to some misclassification, suggesting potential improvements in the classifier’s semantic understanding through expanded training on nuanced language patterns.

\begin{figure}[htbp]
\centering
\includegraphics[width=\columnwidth]{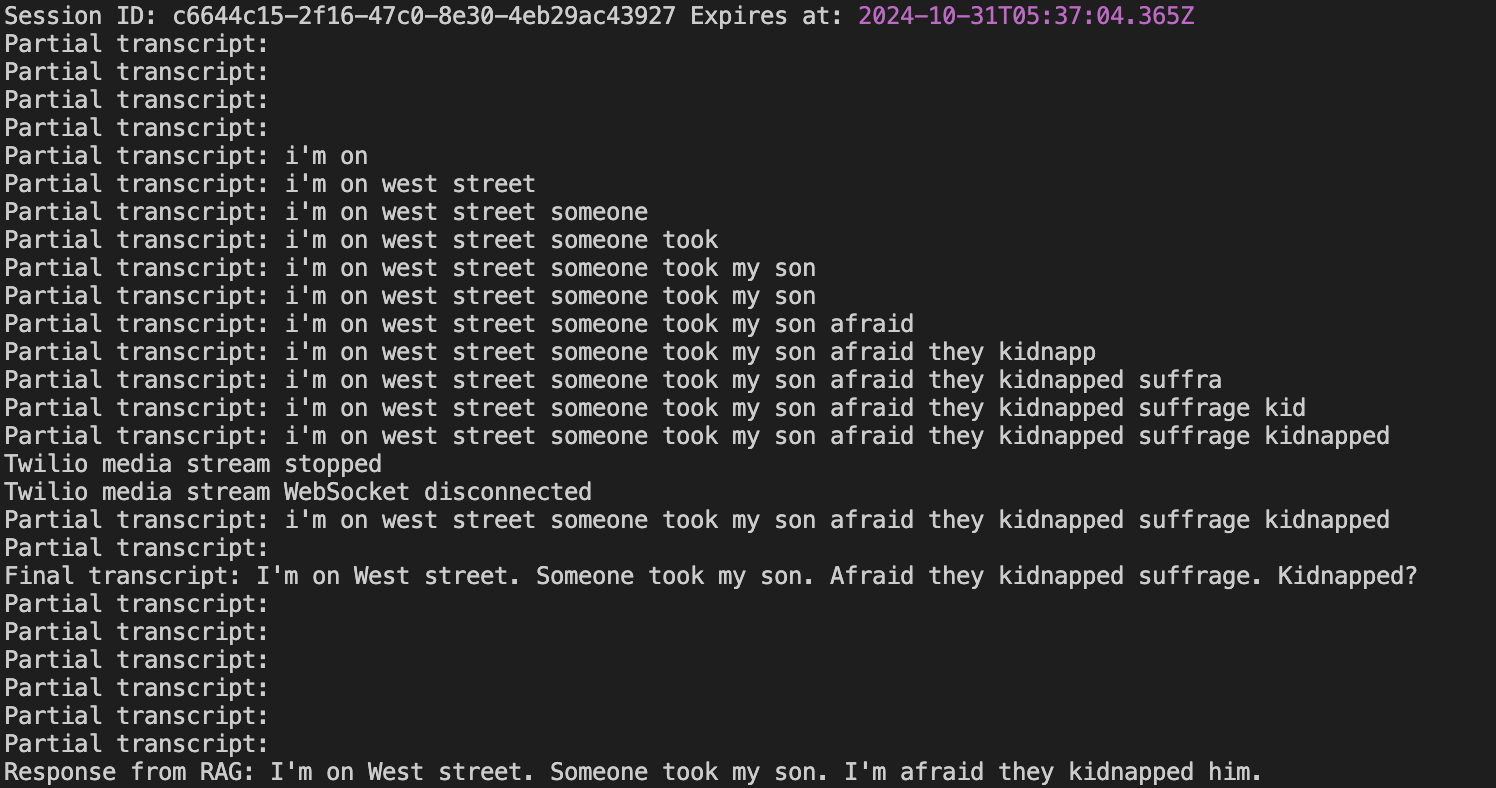} % Adjust the width to \columnwidth
\caption{Real time transcript using RAG: Sample Output}
\label{fig:sampleoutput}
\end{figure}

Fig.~\ref{fig:sampleoutput} shows a console output of a real-time transcription system processing with Twilio media stream a communication network. The figure showcases partial transcripts that update dynamically as the caller speaks, despite challenges such as packet loss and signal degradation. The system’s ability to generate coherent and relevant responses from partial, intermittent transcripts highlights its robustness in VoIP settings, where network quality can vary. This resilience is further supported by the RAG model’s integration, which enhances the accuracy and relevance of final transcripts, ultimately improving response times and the precision of the information relayed to emergency responders.

\section{Conclusion and Future Work}
This study demonstrates the effectiveness of an AI-based system for real-time transcription, intent prediction, and severity classification of emergency communications. By leveraging the LLM, the system successfully interprets partial or unclear transcripts, providing contextually relevant and enriched predictions that improve understanding and prioritization of emergencies. The low BLEU scores across various scenarios indicate that the system effectively goes beyond basic transcription to capture essential contextual details. At the same time, the high conceptual precision confirms its ability to interpret the intent of each communication correctly. The rule-based severity classification also ensures that high-risk situations are flagged and prioritized promptly, reducing the response time for critical emergencies. The system's effectiveness is impacted by common VoIP challenges, such as packet loss, limited bandwidth, and varying audio quality. These factors can degrade audio input, resulting in incomplete or distorted transcripts, which, in turn, can affect the accuracy of intent prediction and severity classification. Despite these challenges, the system shows resilience in interpreting and generating contextually relevant responses, as demonstrated by high conceptual precision even under degraded conditions. This resilience is crucial in emergency settings, where reliable communication may be compromised by network instability or poor audio quality. Overall, this system provides a comprehensive solution for enhancing emergency call management, showcasing the potential of AI to improve efficiency, accuracy, and decision-making in real-world crises.
Although the existing system can manage English-language transcripts efficiently, future versions could broaden their scope to include support for multiple languages. This expansion would enable the system to function effectively in various linguistic settings, ensuring precise transcription and intent prediction for speakers across different languages.
Future enhancements could involve training or refining the LLM to understand better more complex language, such as slang, abbreviations, and informal expressions that may arise during emergencies. 
%This would help minimize errors in classification and boost the accuracy of intent prediction.
By integrating audio analysis tools that can recognize stress or urgency in a caller's voice, the system could gain additional insights for severity classification. 
%This enhancement would further improve the system's ability to prioritize cases, mainly when verbal indicators may not fully reflect the urgency of an emergency.
Future efforts could focus on extensive testing and implementation in collaboration with emergency response organizations to assess the system's effectiveness in real-world scenarios. This would yield valuable information regarding the system's performance in real-time situations, highlight potential improvement areas, and refine algorithms based on actual data.
%Introducing a feedback system where operators can share their insights on the system’s predictions could facilitate ongoing learning and enhancement. This would allow the system to evolve, learning from corrections to improve future predictions and classifications.
%Currently, the system employs a rule-based method for severity classification. Future initiatives could investigate deep learning models capable of identifying complex patterns and contexts for more precise severity classification, reducing dependence on fixed keywords and increasing adaptability.

\bibliographystyle{IEEEtran}
\bibliography{IEEEabrv,References/mybib}

% Generated by IEEEtran.bst, version: 1.14 (2015/08/26)
\begin{thebibliography}{10}
\providecommand{\url}[1]{#1}
\csname url@samestyle\endcsname
\providecommand{\newblock}{\relax}
\providecommand{\bibinfo}[2]{#2}
\providecommand{\BIBentrySTDinterwordspacing}{\spaceskip=0pt\relax}
\providecommand{\BIBentryALTinterwordstretchfactor}{4}
\providecommand{\BIBentryALTinterwordspacing}{\spaceskip=\fontdimen2\font plus
\BIBentryALTinterwordstretchfactor\fontdimen3\font minus \fontdimen4\font\relax}
\providecommand{\BIBforeignlanguage}[2]{{%
\expandafter\ifx\csname l@#1\endcsname\relax
\typeout{** WARNING: IEEEtran.bst: No hyphenation pattern has been}%
\typeout{** loaded for the language `#1'. Using the pattern for}%
\typeout{** the default language instead.}%
\else
\language=\csname l@#1\endcsname
\fi
#2}}
\providecommand{\BIBdecl}{\relax}
\BIBdecl

\bibitem{Venkateshperumal2024RAG}
D.~V, ``Emergency llm,'' \url{https://github.com/danushv/LLM_EmergencyService}, 2024.

\bibitem{nena_stats}
\BIBentryALTinterwordspacing
NENA, ``9-1-1 statistics,'' 2024, accessed: 2024-10-05. [Online]. Available: \url{https://perma.cc/RS7Y-SZEJ}
\BIBentrySTDinterwordspacing

\bibitem{vriezekolk2016assessing}
E.~Vriezekolk, ``Assessing telecommunication service availability risks for crisis organisations,'' 2016.

\bibitem{911Survey}
{National 911 Program}, ``Survey: More than half of 911 centers face staffing crisis,'' \url{https://www.911.gov/newsletters/issue-14/survey-more-than-half-of-911-centers-face-staffing-crisis/}, 2022, accessed: 2024-10-05.

\bibitem{IAEDPressRelease}
{International Academies of Emergency Dispatch}, ``Press release,'' \url{https://www.emergencydispatch.org/in-the-news/press-releases/64632133-7f9f-4d77-8013-d59c445fdb88}, 2022, accessed: 2024-10-05.

\bibitem{neusteter2019911}
S.~R. Neusteter, M.~Mapolski, M.~Khogali, and M.~O'Toole, ``The 911 call processing system: A review of the literature as it relates to policing,'' 2019.

\bibitem{alshammari2024impact}
\BIBentryALTinterwordspacing
F.~Alshammari, S.~Alanazi, and M.~Alqhtani, ``The impact of response time of emergency medical services on fatality rates in an urban environment,'' \emph{International Journal of Medical Informatics and Research Management}, 2024. [Online]. Available: \url{https://ijmirm.com/index.php/ijmirm/article/view/126}
\BIBentrySTDinterwordspacing

\bibitem{brodsky1990emergency}
\BIBentryALTinterwordspacing
H.~Brodsky, ``Emergency medical service rescue time in fatal road accidents,'' in \emph{Transportation Research Record}, 1990. [Online]. Available: \url{https://onlinepubs.trb.org/Onlinepubs/trr/1990/1270/1270-011.pdf}
\BIBentrySTDinterwordspacing

\bibitem{wells2020emergency}
\BIBentryALTinterwordspacing
A.~L. Wells, ``Emergency communications: A study of 911 call failures and resulting fatalities,'' Ph.D. dissertation, Capella University, 2020, accessed: November 5, 2024. [Online]. Available: \url{https://www.proquest.com/docview/2509598984?pq-origsite=gscholar&fromopenview=true&sourcetype=Dissertations%20&%20Theses}
\BIBentrySTDinterwordspacing

\bibitem{fcc_voip_911}
\BIBentryALTinterwordspacing
{Federal Communications Commission}, ``{VoIP and 911 Service},'' 2024, accessed: 2024-11-07. [Online]. Available: \url{https://www.fcc.gov/consumers/guides/voip-and-911-service}
\BIBentrySTDinterwordspacing

\bibitem{james2004implementing}
J.~H. James, B.~Chen, and L.~Garrison, ``Implementing voip: a voice transmission performance progress report,'' \emph{IEEE Communications Magazine}, vol.~42, no.~7, pp. 36--41, 2004.

\bibitem{mansfield2009computer}
K.~C. Mansfield and J.~L. Antonakos, \emph{Computer Networking for LANS to WANS: Hardware, Software and Security}.\hskip 1em plus 0.5em minus 0.4em\relax Delmar Learning, 2009.

\bibitem{sun2006voice}
L.~Sun and E.~C. Ifeachor, ``Voice quality prediction models and their application in voip networks,'' \emph{IEEE Transactions on Multimedia}, vol.~8, no.~4, pp. 809--820, 2006.

\bibitem{boutremans2002impact}
C.~Boutremans, G.~Iannaccone, and C.~Diot, ``Impact of link failures on voip performance,'' in \emph{Proceedings of the 12th International Workshop on Network and Operating Systems Support for Digital Audio and Video}, 2002, pp. 63--71.

\bibitem{manousos2005voice}
M.~Manousos, S.~Apostolacos, I.~Grammatikakis, D.~Mexis, D.~Kagklis, and E.~Sykas, ``Voice-quality monitoring and control for voip,'' \emph{IEEE Internet Computing}, vol.~9, no.~4, pp. 35--42, 2005.

\bibitem{burroughs2017three}
J.~E. Burroughs, ``Three factors leading to failed communications in emergency situations,'' Ph.D. dissertation, Walden University, 2017.

\bibitem{blackwell2002response}
T.~H. Blackwell and J.~S. Kaufman, ``Response time effectiveness: comparison of response time and survival in an urban emergency medical services system,'' \emph{Academic Emergency Medicine}, vol.~9, no.~4, pp. 288--295, 2002.

\bibitem{acuna2020ambulance}
J.~A. Acuna, J.~L. Zayas-Castro, and H.~Charkhgard, ``Ambulance allocation optimization model for the overcrowding problem in us emergency departments: A case study in florida,'' \emph{Socio-Economic Planning Sciences}, vol.~71, p. 100747, 2020.

\bibitem{eslami2024covid}
M.~Eslami~Jahromi, H.~Ayatollahi, and A.~Ebrazeh, ``Covid-19 hotlines, helplines and call centers: a systematic review of characteristics, challenges and lessons learned,'' \emph{BMC Public Health}, vol.~24, no.~1, p. 1191, 2024.

\bibitem{meischke2010emergency}
H.~Meischke, D.~Chavez, S.~Bradley, T.~Rea, and M.~Eisenberg, ``Emergency communications with limited-english-proficiency populations,'' \emph{Prehospital Emergency Care}, vol.~14, no.~2, pp. 265--271, 2010.

\bibitem{carroll2013serving}
L.~N. Carroll, R.~E. Calhoun, C.~C. Subido, I.~S. Painter, and H.~W. Meischke, ``Serving limited english proficient callers: a survey of 9-1-1 police telecommunicators,'' \emph{Prehospital and disaster medicine}, vol.~28, no.~3, pp. 286--291, 2013.

\bibitem{poranen2024human}
A.~Poranen, A.~Kouvonen, and H.~Nordquist, ``Human errors in emergency medical services: a qualitative analysis of contributing factors,'' \emph{Scandinavian Journal of Trauma, Resuscitation and Emergency Medicine}, vol.~32, no.~1, p.~78, 2024.

\bibitem{blandford2004situation}
A.~Blandford and B.~W. Wong, ``Situation awareness in emergency medical dispatch,'' \emph{International journal of human-computer studies}, vol.~61, no.~4, pp. 421--452, 2004.

\bibitem{powers2023using}
C.~J. Powers, A.~Devaraj, K.~Ashqeen, A.~Dontula, A.~Joshi, J.~Shenoy, and D.~Murthy, ``Using artificial intelligence to identify emergency messages on social media during a natural disaster: A deep learning approach,'' \emph{International Journal of Information Management Data Insights}, vol.~3, no.~1, p. 100164, 2023.

\bibitem{sufi2022automated}
F.~K. Sufi and I.~Khalil, ``Automated disaster monitoring from social media posts using ai-based location intelligence and sentiment analysis,'' \emph{IEEE Transactions on Computational Social Systems}, 2022.

\bibitem{madichetty2021stacked}
S.~Madichetty, ``A stacked convolutional neural network for detecting the resource tweets during a disaster,'' \emph{Multimedia tools and applications}, vol.~80, no.~3, pp. 3927--3949, 2021.

\bibitem{otal2024llm}
H.~T. Otal, E.~Stern, and M.~A. Canbaz, ``Llm-assisted crisis management: Building advanced llm platforms for effective emergency response and public collaboration,'' in \emph{2024 IEEE Conference on Artificial Intelligence (CAI)}.\hskip 1em plus 0.5em minus 0.4em\relax IEEE, 2024, pp. 851--859.

\bibitem{chin2021early}
K.-C. Chin, T.-C. Hsieh, W.-C. Chiang, Y.-C. Chien, J.-T. Sun, H.-Y. Lin, M.-J. Hsieh, C.-W. Yang, A.~Y. Chen, and M.~H.-M. Ma, ``Early recognition of a caller’s emotion in out-of-hospital cardiac arrest dispatching: An artificial intelligence approach,'' \emph{Resuscitation}, vol. 167, pp. 144--150, 2021.

\bibitem{blushtein2020identifying}
O.~Blushtein, M.~Siman-Tov, and R.~Magnezi, ``Identifying and minimizing abuse of emergency call center services through technology,'' \emph{The American Journal of Emergency Medicine}, vol.~38, no.~5, pp. 916--919, 2020.

\bibitem{ma2024leveraging}
Z.~Ma, W.~Wu, Z.~Zheng, Y.~Guo, Q.~Chen, S.~Zhang, and X.~Chen, ``Leveraging speech ptm, text llm, and emotional tts for speech emotion recognition,'' in \emph{ICASSP 2024-2024 IEEE International Conference on Acoustics, Speech and Signal Processing (ICASSP)}.\hskip 1em plus 0.5em minus 0.4em\relax IEEE, 2024, pp. 11\,146--11\,150.

\bibitem{huang2014speech}
Z.~Huang, M.~Dong, Q.~Mao, and Y.~Zhan, ``Speech emotion recognition using cnn,'' in \emph{Proceedings of the 22nd ACM international conference on Multimedia}, 2014, pp. 801--804.

\bibitem{liu2018speech}
Z.-T. Liu, Q.~Xie, M.~Wu, W.-H. Cao, Y.~Mei, and J.-W. Mao, ``Speech emotion recognition based on an improved brain emotion learning model,'' \emph{Neurocomputing}, vol. 309, pp. 145--156, 2018.

\bibitem{gogalniceanu2021capturing}
P.~Gogalniceanu, J.~Olsburgh, I.~Loukopoulos, N.~Sevdalis, and N.~Mamode, ``Capturing the crisis ‘golden moment’--a leadership opportunity for overcoming institutional inertia in safety-critical situations,'' \emph{American Journal of Surgery}, vol. 221, no.~3, p. 622, 2021.

\bibitem{handschu2003emergency}
R.~Handschu, R.~Poppe, J.~Rau{\ss}, B.~Neund{\"o}rfer, and F.~Erbguth, ``Emergency calls in acute stroke,'' \emph{Stroke}, vol.~34, no.~4, pp. 1005--1009, 2003.

\bibitem{lee2021impact}
E.-J. Lee, S.~J. Kim, J.~Bae, E.~J. Lee, O.~D. Kwon, H.-Y. Jeong, Y.~Kim, and H.-B. Jeong, ``Impact of onset-to-door time on outcomes and factors associated with late hospital arrival in patients with acute ischemic stroke,'' \emph{PLoS One}, vol.~16, no.~3, p. e0247829, 2021.

\bibitem{spikecodes_911_call_transcripts}
\BIBentryALTinterwordspacing
Spikecodes, ``911 call transcripts,'' 2023, accessed: 2024-08-27. [Online]. Available: \url{https://huggingface.co/datasets/spikecodes/911-call-transcripts}
\BIBentrySTDinterwordspacing

\end{thebibliography}
% Generated by IEEEtran.bst, version: 1.14 (2015/08/26)

\end{document}